\documentclass[10pt,journal,compsoc]{IEEEtran}

\ifCLASSOPTIONcompsoc
  \usepackage[nocompress]{cite}
\else
  \usepackage{cite}
\fi

\usepackage{framed}
\usepackage{color}
\ifCLASSINFOpdf
   \usepackage[pdftex]{graphicx}
   \DeclareGraphicsExtensions{.pdf,.jpeg,.png}
\else
 
\fi

\usepackage{amsmath}
\usepackage{textcomp}                 
\usepackage{array}
\usepackage{url}
\usepackage{enumitem}
\begin{document}

\title{EpiMob: Interactive Visual Analytics of Citywide Human Mobility Restrictions for Epidemic Control}

\author{Chuang~Yang*, 
        Zhiwen~Zhang*,
        Zipei~Fan*,
        Renhe~Jiang$\dagger$,
        Quanjun~Chen,
        Xuan~Song$\dagger$,
        and~Ryosuke~Shibasaki
\IEEEcompsocitemizethanks{
\IEEEcompsocthanksitem * Equal contribution; $\dagger$ Corresponding Author.
\IEEEcompsocthanksitem C. Yang, Z. Zhang, Z. Fan, R. Jiang, Q. Chen, X. Song, and R. Shibasaki are with the Center for Spatial Information Science, The University of Tokyo. \protect 
E-mail: \{fanzipei,chen1990\}@iis.u-tokyo.ac.jp, \{chuang.yang,zhangzhiwen,jiangrh,songxuan,shiba\}@csis.u-tokyo.ac.jp.

\IEEEcompsocthanksitem Z. Fan, R. Jiang, Q. Chen, X. Song, and R. Shibasaki are also with SUSTech-UTokyo Joint Research Center on Super Smart City, Southern University of Science and Technology.
}
\thanks{Manuscript received xx xxx, 202x; revised xx xxx, 202x.}}

\markboth{Journal of \LaTeX\ Class Files,~Vol.~xx, No.~xx, xxx~202x}%
{Shell \MakeLowercase{\textit{et al.}}: Bare Demo of IEEEtran.cls for Computer Society Journals}

\IEEEtitleabstractindextext{%
\begin{abstract}
The outbreak of coronavirus disease (COVID-19) has swept across more than 180 countries and territories since late January 2020. As a worldwide emergency response, governments have implemented various measures and policies, such as self-quarantine, travel restrictions, work from home, and regional lockdown, to control the spread of the epidemic. These countermeasures seek to restrict human mobility because COVID-19 is a highly contagious disease that is spread by human-to-human transmission. Medical experts and policymakers have expressed the urgency to effectively evaluate the outcome of human restriction policies with the aid of big data and information technology. Thus, based on big human mobility data and city POI data, an interactive visual analytics system called Epidemic Mobility (EpiMob) was designed in this study. The system interactively simulates the changes in human mobility and infection status in response to the implementation of a certain restriction policy or a combination of policies (e.g., regional lockdown, telecommuting, screening). Users can conveniently designate the spatial and temporal ranges for different mobility restriction policies. Then, the results reflecting the infection situation under different policies are dynamically displayed and can be flexibly compared and analyzed in depth. Multiple case studies consisting of interviews with domain experts were conducted in the largest metropolitan area of Japan (i.e., Greater Tokyo Area) to demonstrate that the system can provide insight into the effects of different human mobility restriction policies for epidemic control, through measurements and comparisons.

\end{abstract}

}

\maketitle

\IEEEdisplaynontitleabstractindextext

\IEEEpeerreviewmaketitle

\IEEEraisesectionheading{\section{Introduction}\label{sec:introduction}}
\IEEEPARstart{C}{oronavirus} disease (COVID-19) has been spreading in more than 180 countries and territories since late January 2020 and has caused significant damage to public health services as well as to the worldwide economy. In response to the COVID-19 emergency, governments have implemented various measures and policies to restrict human mobility, such as self-quarantine, travel restrictions, working from home, and regional lockdown to contain the rapid spread of the pandemic \cite{Hale2021AGP}. To formulate rational and scientific measures, exploring the effectiveness of these mobility intervention policies has become a significant and urgent issue. Intense research efforts have been expended in this regard. For example, from the modeling perspective, the potential effect of implementing a series of travel restriction policies in China \cite{yang2020modified} and Italy \cite{giordano2020modelling} has been estimated.

Analysts and decision-makers, however, often face a large state space of policies during decision-making (i.e., when and where for what policy). To simplify the policy search process and achieve effective and efficient decision-making outcomes, some efforts centered on interactive simulation and analysis of interventions have been proposed \cite{Eichner2007TheIP,Broeck2011TheGC,afzal2011visual,Afzal2020AVA,Yez2017PandemCapDS}. Most of these efforts merely offered the opportunity to manipulate the simulation at county and higher spatial scales, simplifying citywide prevention and control scenarios (i.e., treating the city as a minimum implementation unit). As the epidemic normalizes, citywide fine-grained epidemic control and prevention could be a more appropriate route for restraining the spread of the infection while maintaining normal livelihoods \cite{Fan2021FinegrainedDR}. A highly versatile simulator that can easily and rapidly simulate and analyze various intracity mobility control strategies would be significant, e.g., supporting high-risk areas' identification and lockdown. To this end, together with domain experts, we designed a visual analytics system for citywide epidemic control scenarios, named EpiMob (\underline{Epi}demic \underline{Mob}ility), capable of interactive simulation and analysis of different mobility policies, to provide decision support to city managers and medical experts. Developing such a system confronts challenges from three aspects.

\begin{figure*}[!t]
	\centering
	\includegraphics[width=0.95\textwidth]{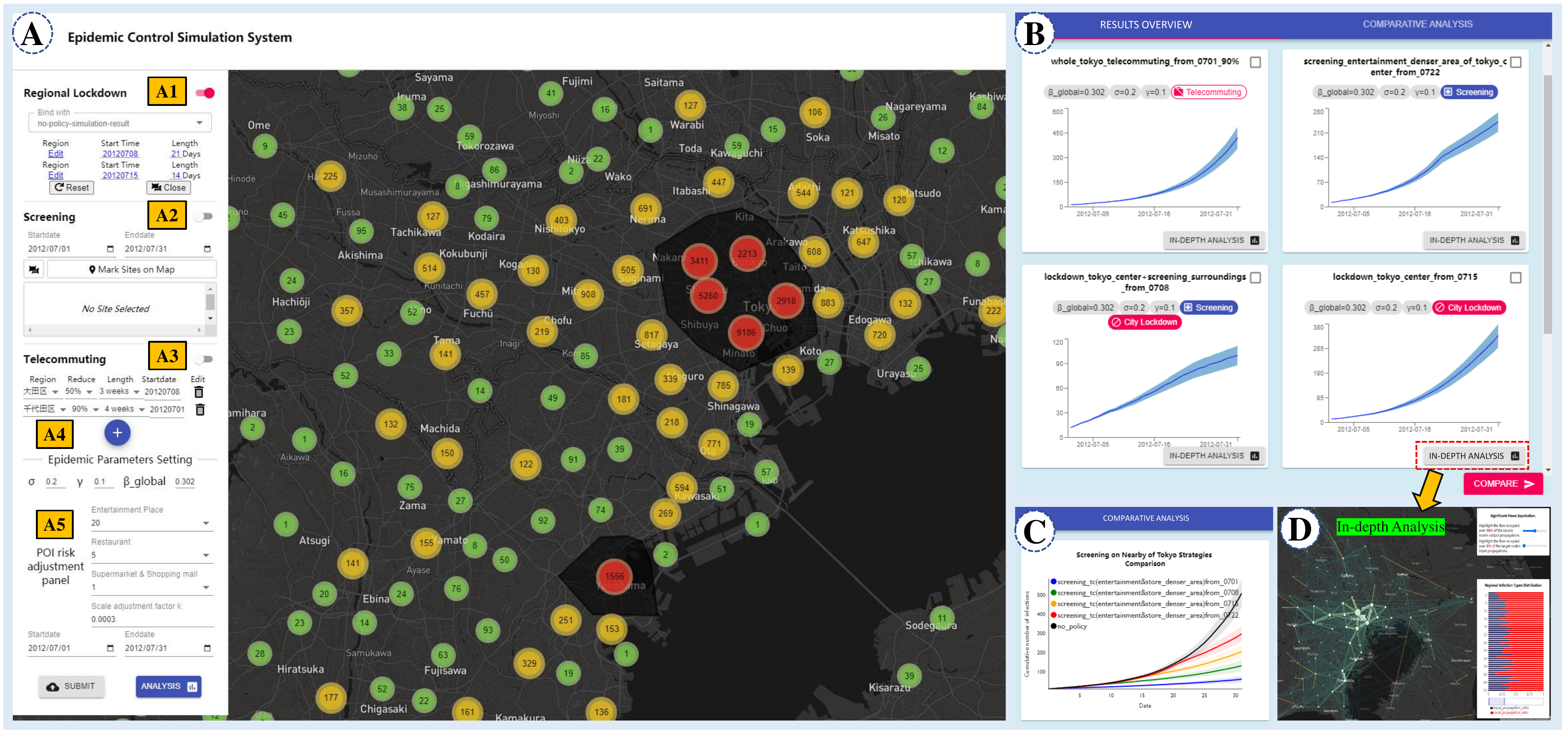} 
	\caption{EpiMob--an interactive visual analytics system for simulating and evaluating the effects of human mobility restriction policies for epidemic control. In Panel A the user is enabled to specify the mobility restriction policies, including A1---``regional lockdown,'' A2---``screening,'' and A3---``telecommuting.'' To adapt to different diseases and different local environments, users can further set the essential epidemic parameters (A4, A5). The results of the analysis on transmission and infection are displayed in the policy result overview panel B. By clicking the button in the bottom right corner of a simulation result, the user can further perform an in-depth analysis of its spatial propagation feature (panel D). In addition, the user can perform a comparative analysis of different policies by selecting them in panel B, and the results are displayed in panel C. 
	\textbf{Note:} There is no content consistency among B and C, the purpose was merely to express the operation logic of comparative analysis.} 
	\label{fig:teaser}
\end{figure*}
\noindent \textbf{A. Uniqueness and complexity of the citywide governance.} Compared to the higher spatial scales (i.e., county to global), the citywide decision-making scenario involves local control policies and requires complex intracity features to be considered, such as regional exposure and diffusion risks. Customized visualization views and interaction logic are required to efficiently perceive critical information and set policies. Existing simulators were not designed for such a decision environment (Section~\ref{sec:rlw_vis}), posing new requirements and challenges to our visualization design.

\noindent \textbf{B. Spatiotemporal heterogeneity of modeling environment}. Intracity environment features, such as regional functions and population fluxes change over time and space, which are essential for citywide fine-grained propagation modeling. However, existing models that consider these are too complex for interactive scenarios \cite{eubank2004modelling,chang2020mobility,Barrett2008EpiSimdemicsAE}. Experts desire to obtain numerous simulation insights as quickly as possible. Thus, a model balancing complexity with computational efficiency is required, presenting challenges to our model design.

\noindent \textbf{C. Diversity and complexity of simulation conditions}. The types, intensities, and spatiotemporal scope of restriction policies are diverse. Decision makers may perform multiple types of policies simultaneously. Moreover, given a policy, individual response behavior varies significantly from person to person. The existing work accommodates these scenarios through direct and sophisticated parameter settings (Section~\ref{sec:rlw_vis}). However, to devise a highly versatile simulator that is easy to use while supporting wide control policies, the user experience needs to be balanced against parametric complexity, which poses challenges to our system design.

For \textbf{Challenge A}, 
we integrate some recent visualization methods/views with an easy-to-use interactive logic to promote the exploration, setting, evaluation, and analytics of citywide movement restriction strategies.
Fig. \ref{fig:teaser} shows the main user interface of our system. It enables users to specify three widely used mobility restriction policies: screening, telecommuting, and regional lockdown. Users can conveniently explore and designate spatiotemporal ranges for different mobility restriction policies (e.g., starting a telecommuting policy for the entire central business district of Tokyo from March 1, 2020.) with the help of informative spatial views and easy-to-use interaction logic (Fig. \ref{fig:teaser}-A). The results of simulating the infection situation are dynamically displayed on the display panel (Fig. \ref{fig:teaser}-B), and the user can further perform comparative analysis (Fig. \ref{fig:teaser}-C) and in-depth exploration (Fig. \ref{fig:teaser}-D). 
A novel trajectory-based epidemic model was proposed for \textbf{Challenge B}. The model is driven by human trajectories and point of interest (POI) data to capture the spatiotemporal heterogeneity of the citywide modeling environment. It can dynamically and continuously perform grid-level fine-grained simulations at a fixed time frequency (e.g., every 5 min), achieving a balance between model complexity and computational efficiency.
For \textbf{Challenge C}, a novel simulation mechanism was devised, following the design principle of \textit{``high cohesion and low coupling.''} When executing a mobility restriction policy, the simulation process is divided into two stages: generating restricted mobility first and then performing the epidemic simulation. 
Such a mechanism resorts to mobility changes to reflect the effects of restriction policies rather than imposing them directly on the model settings, enhancing the usability and extensibility of our system. 
In this work, a \textit{``trajectory replacement''} restriction generation strategy was proposed to produce the restricted mobility, simplifying the sophisticated parameter settings of the policies.

By performing multiple case studies of the largest metropolitan area in Japan (i.e., Greater Tokyo Area) and domain expert interviews, it is demonstrated that our system can provide illustrative insight into exploring and analyzing the effects of different human mobility restriction policies for epidemic control.
To the best of our knowledge, EpiMob is the first interactive visual analytics system that can provide epidemic control policy simulation at fine-grained spatiotemporal granularity by utilizing citywide human mobility data and city POI data. The major contributions of our study are summarized as follows:
\begin{itemize}[leftmargin=*]
\item A visual analytics solution integrating some recent visualization methods with an easy-to-use interactive logic is proposed, to help the end user interactively explore, set, and analyze different mobility restriction policies and corresponding quantitative effects efficiently.
\item A ``trajectory replacement`` strategy is designed to accommodate different settings on human mobility restrictions. Based on this strategy, an online web system was implemented with a well-modularized architecture.
\item A novel trajectory-based epidemic model was proposed to simulate the fine-grained spreading of an epidemic based on real-world human trajectory data and city POI data.
\item The proposed system was evaluated by conducting multiple case studies as well as interviews with domain experts, demonstrating the superior performance, functionality, and usability of our system. 
\end{itemize}

\section{Related Work} \label{sec:rlw} 
\subsection{Evaluation of Epidemic Control Measures} \label{sec:rlw_eva} 
Evaluating the effectiveness of intervention measures for epidemic control is a significant and extensively studied issue. Many research efforts have been conducted, including pharmaceutical interventions~\cite{mukandavire2020quantifying,bartsch2020vaccine,eubank2004modelling} (e.g., vaccination), and nonpharmaceutical interventions~\cite{silva2020covid,Lai2020EffectON,tian2020investigation,giordano2020modelling,chinazzi2020effect,yang2020modified,Gatto2020SpreadAD,eubank2004modelling,chang2020mobility} (e.g., social distance and mobility restriction). Our study falls into the second category and concentrates on mobility restrictions. Utilizing real-world mobility data, researchers have analyzed the effectiveness of mobility restriction policies at the city level \cite{eubank2004modelling,chang2020mobility}, domestic level~\cite{tian2020investigation,Lai2020EffectON,yang2020modified,chinazzi2020effect,Gatto2020SpreadAD,giordano2020modelling}, and international level~\cite{chinazzi2020effect}.
The restriction types encompass regional travel restrictions, lockdowns, and quarantine. Most of these studies evaluated the intervention after its implementation. For future trend prediction, using the city-to-city in-out flow data, a modified SEIR and AI prediction model was proposed to predict the COVID-19 epidemic peaks and sizes under various travel restrictions and social intervention policies \cite{yang2020modified}. Giordano \textit{et al.} predicted the epidemic evolution in Italy under different lockdown intensities \cite{giordano2020modelling}.
The main differences between these studies and ours are as follows: (i) They specify policies through direct parameter settings, for example, by manually adjusting migration rates to simulate travel restriction policies \cite{yang2020modified}. This study focuses on interactively setting and simulating policies, highlights when and where for what policies. (ii) This work is geared towards citywide epidemic control, which introduces a finer granularity into policy implementation. In comparison, most of the above works are designed with coarser granularity.

\subsection{Epidemic Analytics and Visualization} \label{sec:rlw_vis} 
The efforts for the epidemic visualization can be summarized under three categories. (A) \textbf{Disease Characteristics}: visualizing disease-related characteristics, such as virus structure~\cite{Nguyen2021ModelingIT}, region-based features~\cite{Dong2020AnIW}, and transmission features~\cite{Baumgartl2021InSO,guo2007visual,dunne2015vorograph}; (B) \textbf{Human Responses Monitoring}: visualizing the human response behavior amidst an epidemic, such as mobility patterns \cite{Gao2020MappingCM,zuo2020interactive} and sentiment~\cite{Naseem2021COVIDSentiAL}; and (C) \textbf{Visualization-assisted Simulation Tools}: Efforts to assist decision-making by visually creating, simulating, and analyzing intervention scenarios \cite{Eichner2007TheIP,Broeck2011TheGC,afzal2011visual,Afzal2020AVA,Yez2017PandemCapDS}. Our work belongs to the third category.
Specifically,
InfluSim \cite{Eichner2007TheIP} provided a purely numerical parameter configuration panel for intervention policy setting.
GLEaMviz \cite{Broeck2011TheGC} supplied a GUI for the users to design dedicated compartmental models for intervention strategies.
Afzal \textit{et al.} proposed a decision history view \cite{afzal2011visual} to compare the effects of time-varying strategy combinations, and an extension of \cite{afzal2011visual} has been proposed for the modeling, simulation, and exploration of the spread of COVID-19 \cite{Afzal2020AVA}.
PandemCap \cite{Yez2017PandemCapDS} provided a series of statistical charts to present and compare the simulation results.
The objective of our study is similar to that of previous studies; however, ours is distinct in the following aspects: (i) None of the studies mentioned above focus on citywide fine-grained transmission control, instead of on higher spatial units \cite{Broeck2011TheGC,Afzal2020AVA,afzal2011visual} or a single area without considering the spatial structure \cite{Eichner2007TheIP,Yez2017PandemCapDS}. 
(ii) The above studies simulate different restriction scenarios by manually specifying the restriction-related model parameters \cite{Eichner2007TheIP,afzal2011visual,Afzal2020AVA,Yez2017PandemCapDS} or visually configuring new models \cite{Broeck2011TheGC}. Our work indirectly reflects policies through the interactive simulation of restricted mobility.
Because real-world human mobility data characterizes the fine-grained mobility dynamics and human behaviors of cities \cite{phithakkitnukoon2010activity}, fewer parameter settings are needed in our model.

\subsection{Citywide Mobility Simulation}
Mobility simulation at the citywide level has been a major challenge in the last decade. The Brinkhoff generator \cite{brinkhoff2002framework} was proposed to generate and simulate network-based trajectories given the road network of a city. MNTG \cite{mokbel2013mntg} extended the Brinkhoff generator \cite{brinkhoff2002framework} to a web-based traffic generator. SUMO \cite{behrisch2011sumo} can simulate human mobility in a large urban area. Multi-agent transport simulation (MATSim) \cite{horni2016multi} is a state-of-the-art solution for citywide traffic simulations in the field of transportation engineering. In addition, data-driven mobility simulation/generation has been proposed for the large urban area \cite{baratchi2014hierarchical,ouyang2018non,kang2020trag,song2015simulator}. Although these studies can be applied to simulate citywide human mobility in normal or emergency situations, the simulation or generation of citywide human mobility under different restriction policies in an epidemic or pandemic situation has not yet been reported. To address this shortcoming, here, a novel replacement-based mobility generation strategy is proposed.

\section{Background}
In this section, we first formulate our problem, then provide the task analysis, and finally describe the data source.
\subsection{Problem Formulation} \label{sec:profom}
Large-scale and high-intensity mobility restrictions (e.g., city lockdown) significantly impact people's livelihoods and the economy. With the normalization of the epidemic, local and fine-grained outbreak control may be a more appropriate approach, necessitating the evaluation and analysis of the effects of different intracity mobility control strategies. By conducting semi-structured interviews with domain experts with a strong background in epidemiology, it was identified that the current citywide simulation method cannot efficiently do so, which can be summarized under two main points: \textbf{P1}. \textit{Tedious modeling process}. Experts need to employ agent-based models and manually build the constraints, which is very time consuming, inflexible, and requires high resource consumption and expertise (e.g., defining human interaction behaviors); \textbf{P2}. \textit{Inefficient policy search process}: There are no unified platforms/simulators designed to explore when-and-where-to-apply-what-policy in a city. Current practices implement manual configuration and extensive statistical analysis for specific policies, combined with static charts/diagrams for visualization.

To this end, we collaborated with three domain experts to pool expertise and build an insightful and user-friendly simulator for decision support in citywide epidemic control scenarios. The simulator is mainly driven by citywide trajectory data, which profiles the detailed, diverse, and complex features of the individual's movement, that is, the time, location, and even semantic information, and is relevant to the spread and control of infectious diseases \cite{Gonzlez2008UnderstandingIH,phithakkitnukoon2010activity}.
EA is an experienced infectious disease specialist who has achieved several results in COVID-19 prevention and treatment.
EB is a proficient researcher in complex networks and has conducted several studies on human mobility networks and modeling infectious diseases, including collaboration with public health researchers to provide suggestions for countermeasures against the spread of infection.
EC is an expert focusing on interdisciplinary research on public health and GIS, especially in spatial epidemiology. 
They all have extensive experience in modeling infectious diseases and have used their expertise to assist decision-making.
All three experts were consulted while designing the epidemic simulation model. We also gathered the visual design requirements from them and iteratively collected feedback during the design process. Moreover, EB also acted as a consultant by advising on the relationship between human mobility and policy restrictions.

\subsection{Workflow Construction} \label{sec:basicreq}
To clarify the working logic of the system, we held a preliminary requirements gathering meeting and constructed a four-stage workflow for efficient citywide policy-making.

\noindent\textbf{Configuration}: At this stage, the experts aim to configure the disease transmission parameters, which is an indispensable requirement for any simulation task. 

\noindent\textbf{Exploration}: At this stage, the experts aim to determine when and where to deploy what policies (\textbf{P2}) with the help of efficient and effective information views. Correspondingly, an easy-to-use interaction logic enabling conveniently simulation setting and launching is necessary.

\noindent\textbf{Simulation}: At this stage, with the model parameters and policy, the simulation mechanism shall be well suited to accept and execute complicated policies (\textbf{P1}). 

\noindent\textbf{Evaluation}: Given the simulation results, the fundamental demands of a simulator are to evaluate the results of a single policy and compare the advantages and disadvantages of different policies to find the best candidates.

\subsection{Task Analysis} \label{sec:taskanalysis}
During the past year, we held a series of virtual meetings with the experts to discuss requirements and collect feedback. The workflow was iteratively improved and the following requirements were derived:

\noindent\textbf{Configuration Stage (C).}

\noindent\textbf{C1: Basic parameter setting of propagation model.}
As analyzed in Section \ref{sec:basicreq}, the disease parameter setting is an integral part of any simulation tool. For COVID-19, the experts required a panel to set the disease parameters, that is, recovery/death rate, incubation period, and in particular, the number of adequate exposures per unit time $\beta$ (a healthy person who has adequate exposure to an infectious person is expected to become infected). In terms of the first two parameters, the experts considered that assigning values would be relatively easy because these parameters are relatively stable in a given area. 
However, for setting $\beta$, the experts commented: \textit{``the value of $\beta$ should vary according to the spatiotemporal position, which is highly related to the functions of the regions. Compared to that at higher spatial granularities, it posed new challenges.''} 
For instance, compared with a forest park, people are more likely to be exposed to others in an entertainment center, which would give rise to a higher $\beta$ value. Thus, to build a reasonable fine-grained epidemic model, the system shall assign a varying spatiotemporal $\beta$ value for each region according to its function.
It is clearly impractical for users to manually set varying $\beta$ for each spatiotemporal unit. Hence, by discussing with experts, it was proposed that $\beta$ be inferred from the regional POI information because it adequately reflects the function of the region. However, treating all POI types equally in one region is illogical, the risk differences among them shall be considered. Both our team and the experts think this is an important topic, but it is beyond the scope of this study. Thus, we determined to supply a risk adjustment panel allowing users to set the risk of different types of POI.

\noindent\textbf{Exploration Stage (E).}

\noindent\textbf{E1: A set of spatial views assists policy exploration.}
To solve \textbf{P2}, a special brainstorming session was held to collect reference information on developing regional mobility control policies, which are summarized as follows:
(i) \textit{infection hotspots}. An infection hotspot means that many people are infected in the area relative to other areas. The ability to efficiently locate these areas is beneficial for epidemic control.
To map the citywide infection hotspots, the user must first acquire the infection locations. However, determining the accurate locations of real infected individuals during an outbreak is difficult. Thus, running the simulation to trace and collect infection locations becomes a common strategy. The experts' current solution for infection hotspot visualization included the heatmap and scatter map, which are simple plots without informative interactive design and lacking in-depth exploration. The experts requested us to design a hotspot view that integrates the interaction function and allows for in-depth analysis.
(ii) \textit{workplaces visited with high-frequency.} Many workplaces have emerged as clusters of infection during COVID-19. Identifying workplaces with high-frequency visitation is important in deciding where to enforce remote working. The approach of the experts in representing workplace distribution was limited to the choropleth map at the municipality level. For finer granularity, data-driven workplace estimation is necessary while they do not have related background. Therefore, they wanted us to provide a view displaying the spatial density distribution of people's workplaces.
(iii) \textit{screening point exploration.} The government often set up screening points in certain areas during COVID-19. In practice, the experts identified these areas by plotting a scatter plot of the POI distributions. Then the locations where certain types of POIs are denser were selected. Specifically, these POIs are either at high risk of exposure, such as entertainment places (e.g., bars, karaokes) and restaurants, or have high human traffic such as stations, shopping malls, and public spaces (e.g., parks, zoos, and attractions). The experts suggested integrating this reference information into the system.

\noindent\textbf{E2: Intuitive spatiotemporal policy setting.}
A specific restriction policy must have concrete spatiotemporal information (i.e., the implementation period and regions).
The three view requirements in \textbf{E1} correspond to three policy types: regional lockdown, telecommuting, and screening point deployment. When the user determines the target region in which to launch a policy, conveying the intent to the system becomes important and necessary. For lockdown, experts want to lock down the area of interest directly on a map; for telecommuting, it is more practical for the government to apply it to administrative districts, which are easier to handle. Moreover, working remotely is not practicable for all occupations. Experts wish to take these into consideration; and for screening, experts want to mark the screening points directly on the map. Besides, specifying the execution period is necessary for all policies.

\noindent\textbf{Simulation Stage (S).}

\noindent\textbf{S1: Simulation model supporting dynamic parameter change.}
As discussed in \textbf{C1}, our system should allow users to assign a dynamic $\beta$ value over regions using POI information. Hence, the epidemic model should accept the POI and the corresponding risk information as inputs. 

\noindent\textbf{S2: An extensible and mixable simulation mechanism.}
(i) Supporting both single and compound policies, because \textit{``the real scenario may be quite complicated, and the government employs many strategies in parallel.''}
(ii) Extensible. It should enable other developers to contribute their newly designed strategies easily, instead of being limited to using a few specific policy types.
(iii) Efficiency. Experts reinforced the deficiencies of the existing models: the agent-based model is slow and complex, and the meta-population model is coarse. For interactive citywide simulation, a balance between model complexity and speed needs to be achieved.

\noindent\textbf{Evaluation Stage (L).}

\noindent \textbf{L1: Display of single control policy result.}
(i) Basic requirements: plotting the infection curve is a basic operation to present the results. Meanwhile, because contagion simulations generally have randomness, confidence intervals (CI) should be considered. 
(ii) Distinguishability: There may be sets of simulations with only slight setting differences (e.g., two lockdown policies intersect to a large extent at the target regions).
(iii) In-depth analysis of results: \textit{``What are the secondary effects of implementing the policy? e.g., the new infection hotspots.''} The infection hotspot view can be equipped to directly perceive the transition of hotspots. \textit{``What roles do the regions perform in the spread? Are there any striking patterns of transmission?}``, a network view is required to observe and explore the spatial propagation feature. 

\noindent \textbf{L2: Comparative analysis of different control policies.}
To find the best policy, it is necessary to compare the effects of different control policies based on performance criteria. An interactive logic as convenient as ``select, compare, and display'' is required.

\subsection{Data Description}
\textbf{City POI Data.}
The Telepoint Pack DB of POI was collected in February 2011 provided by ZENRIN DataCom Co., Ltd \cite{poi}. In the original database, each record is a registered landline telephone number with its coordinates (latitude and longitude) and industry category information included. We treated each each ``telepoint'' as a specific POI. All POIs were classified into 39 categories. The total number of POIs in the Greater Tokyo Area was 1,418,563. 

\noindent \textbf{Human Mobility Data.}
In this work, a GPS log dataset anonymously collected from approximately 1.6 million real mobile phone users in Japan over a three-year period (from August 1, 2010, to July 31, 2013) was picked for epidemic simulation. It contains approximately 30 billion GPS records, covering approximately 1\% of the real-world population. The data were collected by a mobile operator (NTT DoCoMo, Inc.) and a private company (ZENRIN DataCom Co., Ltd.) with the consent of the mobile phone users. It was collectively and statistically processed to conceal private information, such as gender and age. By default, the positioning function on the user mobile phones was activated every 5 min. However, data acquisition is affected by several factors, such as signal loss and low battery power. In addition, when a mobile phone user stops at a location, the positioning function is automatically turned off to save power. Furthermore, the age distribution of the dataset skews slightly towards younger users because the young prefer to use a mobile phone with a positioning function compared to users from other age groups (e.g., the elderly). The representativeness of our dataset was verified by previous work \cite{horanont2013databias}, in which the quality of the dataset was evaluated. The home location of each user in the dataset was identified, and the spatial distribution was compared with that of census data on a 1km *1km grid. The following linear correlation was found:
\begin{equation}
    N_{GPS} = 0.0063048*N_{census}+0.73551, R^2 = 0.79222
\end{equation}
where $N_{GPS}$ is the estimated population size, $N_{census}$ is the population size provided by the census data, and $R^2$ is the coefficient of determination.
In this study, the Greater Tokyo Area (including Tokyo Metropolis and the prefectures of Kanagawa, Chiba, and Saitama) was selected as the target area. The user is picked to the experimental data if $80\%$ of its trajectory points are located in the target area. As a result, 145,507 user trajectories were obtained.

\section{Model}
In this section, some basic concepts are initially introduced, followed by the simulation mechanism, and finally, the epidemic model.

\noindent\textbf{Human Trajectory:} The human trajectory collected for an individual essentially comprises a 3-tuple sequence: ($timestamp$, $latitude$, $longitude$), which can indicate a person's location according to a captured timestamp. It can be further denoted as a sequence of ($t$, $l$)-pairs to which the user ID $uid$ is attached by simplifying $timestamp$ as $t$ and ($latitude$, $longitude$) as $l$. 
\begin{equation}
traj = \{uid, (t_1,l_1), (t_2,l_2), ..., (t_n,l_n)\}
\end{equation}

\noindent\textbf{Citywide Human Mobility:} Citywide human mobility $\Gamma$ refers to a large group of user trajectories in a given urban area. Given a user ID $uid$, the personal trajectory $\Gamma_{uid}$ is retrieved from $\Gamma$ as $\{(t_1,l_1), (t_2,l_2), ..., (t_n,l_n)\}$.

\noindent\textbf{Grid-mapped Interpolated Human Trajectory:} Raw human trajectory data are not usually sampled at a constant rate. After applying the typical preprocessing method proposed in \cite{jiang2018deepurbanmomentum,jiang2018deep}, the interpolated human trajectory is obtained with a constant sampling rate $\Delta{\tau}$ as follows:
\begin{equation}
    \Gamma_{uid} = \{(t_1, l_1), ..., (t_n,l_n)\}, \forall i \in \left[1,n\right), |t_{i+1} - t_i| = \Delta{\tau}
\end{equation}
where $\Delta{\tau}$ was set to five minutes in our study. Furthermore, the interpolated human trajectory is mapped onto the mesh grid as follows:
\begin{equation}
     \Gamma_{uid} = \{(t_1,g_1), ..., (t_n,g_n)\}, \forall i \in \left[1,n\right], l_i \in g_i
\end{equation}
A third-party geospatial coding library H3\cite{brodsky2018h3} was utilized to complete the grid mapping. H3 divides the earth into seamless spliced hexagons with different spatial resolutions. The selected resolution was level 8 (grid size is approximately 0.737$km^2$), which is considered an appropriate size to balance granularity and speed.

\noindent\textbf{Trajectory-Based Epidemic Simulation:} 
An epidemic can be simulated using the trajectory-based epidemic model $\mathcal{F}_{EPI}$ as follows:
\begin{equation}
E_{sim} = \mathcal{F}_{EPI} (\Gamma \;;\; \Theta)
\end{equation}
where $\Gamma$ is the given citywide human mobility; $\Theta$ refers to the parameters of the infectious disease; and $E_{sim}$ denotes the simulation result. Every time a set of $(\Gamma,\Theta)$ is input into the epidemic model, the epidemic simulation runs anew.

\noindent\textbf{Mobility Restriction Policy:} This study focuses on evaluating the effects of the following restriction policies:
\begin{itemize} [leftmargin=*]
\item \textit{Screening} refers to setting up an infection screening point at a specific location, such as the roadside or a station, to screen whether a person is infectious. 
\item \textit{Telecommuting} is a corporate policy that allows employees to work from home instead of commuting to the office.
\item \textit{Regional Lockdown} is a government policy that implements mandatory geographic quarantine to all citizens living in a specific region (city or ward).
\end{itemize}

\subsection{Two-stage Epidemic Simulation}
To satisfy requirement \textbf{S2}, a two-stage simulation mechanism was devised:

\noindent\textbf{Stage 1-Restricted Mobility Generation:} Given a mobility restriction policy or a combination of several policies $\Phi$, citywide human mobility forcibly changes owing to the given $\Phi$. In this study, a mobility replacement model denoted as $\mathcal{F}_{MOB}$ was utilized to generate the restricted human mobility $\Gamma'$ w.r.t $\Phi$ as follows:
\begin{equation}
\Gamma' = \mathcal{F}_{MOB} (\Gamma \;;\; \Phi)
\end{equation}

\noindent\textbf{Stage 2-Epidemic Simulation with Restricted Mobility:} Given the restricted citywide human mobility $\Gamma'$ w.r.t $\Phi$ and disease transmission parameters $\Theta'$, simulation of the epidemic for the restriction policy settings $E'_{sim}$ can be implemented as follows: 
\begin{equation}
E'_{sim} = \mathcal{F}_{EPI} (\Gamma' \;;\; \Theta')
\end{equation}
This reflects the extensibility of the simulation mechanism. Developers can design new restriction strategies at Stage 1 and use other epidemic models at Stage 2. 

\subsection{Trajectory-Based Epidemic Model}  \label{sec:core_model}
Fundamental compartmental models in epidemiology have been widely applied to predict infectious diseases transmitted from humans to humans, such as measles, mumps, and rubella.
To simulate the contagion process at the grid level, the conventional SEIR model \cite{SEIR} was extended to a grid-based model. Given the grid-mapped interpolated human trajectory of one city and a set of disease parameters, the new model can dynamically and continuously perform fine-grained simulations at a fixed frequency at the grid level.

\subsubsection{Extended SEIR model}
The extended conventional SEIR model is as follows:
\begin{equation}
\begin{aligned}
    \label{eq:sir}
    \frac{dS_{g,\,t}}{dt} &= -\beta_{g,\,t} \frac{S_{g,\,t}I_{g,\,t}}{N_{g,\,t}} \\
    \frac{dE_{g,\,t}}{dt} &= \beta_{g,\,t} \frac{S_{g,\,t}I_{g,\,t}}{N_{g,\,t}} - \sigma E_{g,\,t}\\
    \frac{dI_{g,\,t}}{dt} &= \sigma E_{g,\,t} - \gamma I_{g,\,t} \\
    \frac{dR_{g,\,t}}{dt} &= \gamma I_{g,\,t} \\
    N_{g,\,t} &= S_{g,\,t}+ E_{g,\,t}+ I_{g,\,t}+ R_{g,\,t} \\
\end{aligned}
\end{equation}
where $S_{g,\,t}$, $E_{g,\,t}$, $I_{g,\,t}$, and $R_{g,\,t}$ are the numbers of susceptible, exposed, infected, and recovered users at grid $g$ at time slot $t$. $ N_{g,\,t}$ denotes the total number of people located in grid $g$ at time slot $t$.
$\sigma$ and $\gamma$ are the incubation rate and recovery/death rate, respectively. $\beta_{g,\,t}$ is the number of adequate exposures per unit time at grid $g$ at time slot $t$, designed to support the dynamic $\beta$ setting (\textbf{S1}) consisting of the base ($\beta_{base}$) and varying parts ($\Delta_{\,g,\,t}$), as shown in Equation \ref{eq:beta_vary}; $\beta_{base}$ is the basis of the change in dynamic $\beta$; $\Delta_{\,g,\,t}$ indicates the variation in $\beta_{base}$ at grid $g$ at time slot $t$ when considering the POI risk information. 
\begin{equation}
\begin{aligned} 
     \beta_{g,\,t} &= \beta_{base} + \Delta_{\,g,\,t}   \label{eq:beta_vary} \\
\end{aligned} 
\end{equation}
\noindent Because there are numerous types of POI scattered in a grid, and over a time interval $t$, some are open for business, and some are closed. The idea is to request experts to assign a risk value to each POI based on their experience. Then, the cumulative risk value of all open POIs at grid $g$ at time slot $t$ is $\Delta_{\,g,\,t}$. However, the cumulative risk value and $\beta$ values are not of the same order of magnitude. Thus, a scale adjustment factor $k$ is introduced to balance two orders of magnitude. Equation \ref{eq:beta_delta} formulates this idea: 
\begin{equation}
\begin{aligned} 
     \Delta_{\,g,\,t} &= k \times R_{g,\,t}
       =  k \times \sum_{i=1}^{n} p_{g,\,t,\,i} \times r_{i} \label{eq:beta_delta}\\
\end{aligned} 
\end{equation}
$R_{g,\,t}$ denotes the cumulative risk value; $p_{g,\,t,\,i}$ indicates the number of open POIs belonging to category $i$; and $r_{i}$ denotes the risk value of each POI for that category. Our model requires users to specify $r_{i}$ and $k $ to obtain $\Delta _{\, g, \, t} $. 

Compared to obtaining $\beta_{base}$, it is relatively easy to estimate a spatiotemporal invariant $\beta_{global}$, on which the researchers have put in considerable effort \cite{linton2020incubation,kuniya2020prediction}. Assume that the following relationship exists between $\beta_{base}$ and $\beta_{global}$:
\begin{equation}
\begin{aligned} 
     \beta_{global} &= \overline{\beta_{g,\,t}} 
                    = \beta_{base} + \overline{\Delta_{\,g,\,t}}  \label{eq:beta_global}\\
\end{aligned} 
\end{equation}
$\overline{\beta_{g,\,t}}$ and $\overline{\Delta_{\,g,\,t}}$ are the average values of $\beta_{g,\,t}$ and $\Delta_{\,g,\,t}$, respectively.
Thus, to obtain $\beta_{base}$, our model requires users to specify $\beta_{global}$.

\subsubsection{Simulation Process}
\begin{figure}[ht]
\centering
\includegraphics[width=0.44\textwidth]{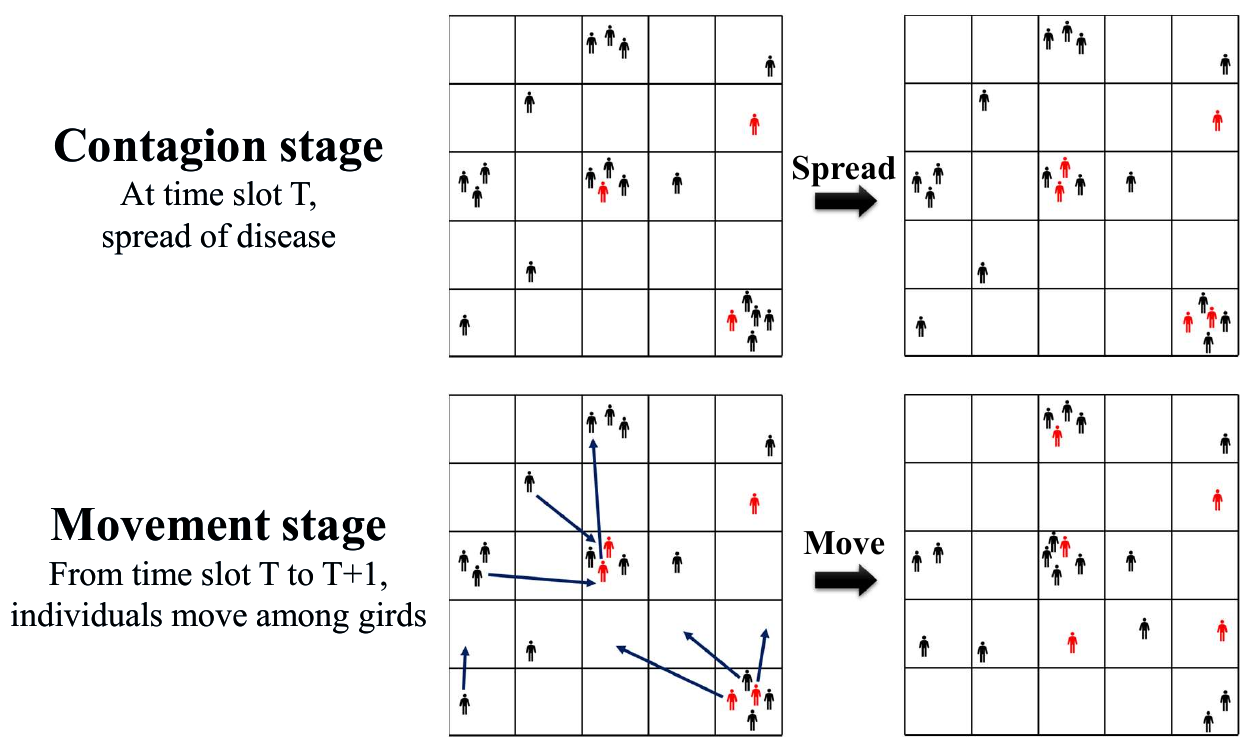}
\caption{Illustration of the simulation process. The contagion stage simulates the spread of the disease at each grid at time slot $T$ and updates the user state. From $T$ to $T+1$, the movement stage updates the user list of each grid in response to user movements. The color of the humanoid marker indicates the user state: red for infected, and black for susceptible. The hexagon is simplified to a cell for drawing.}
\label{fig:epi_model}
\end{figure}
An iterative simulation process called ``\textit{contagion + movement}'' is proposed (Fig. \ref{fig:epi_model}). In the \textit{contagion} stage, it is assumed that each person has a certain probability of being infected by others inside the same grid at the same time slot. In the \textit{movement} stage, the user movements update the user lists of grids and introduce the disease to other grids. By iterating the process over the target time range, each user's state change is traced and the infection events are collected as simulation results. During the simulation, once a user's health state changes from susceptible to exposed, the corresponding infection time and location are recorded as an infection event. To start the iterative simulation process, $I_{0}$ initially infected individuals was selected by random sampling, and hence, their user state was updated to ``infected.'' By default, $I_{0}$ was set to 10 in this study.

Unlike the classical SEIR model, the proposed model exhibits randomness. Given $(\Gamma,\Theta)$, the simulation result $E_{sim}$ is unstable with respect to the total number of infection events and their occurrence times and locations. Therefore, multiple simulations are necessary. The number of repetitions $m$ was set to 100 in this study. 
In addition, parallel computing was applied for the repetition simulations to ensure computational efficiency. For more details on the model, please see the supplementary materials.

\subsection{Replacement-Based Restricted Mobility Model} 
When a user is affected by a mobility restriction policy, the mobility behavior is influenced accordingly. For example, a user frequents a shopping mall every weekend. The mall is locked on a particular day, and the user's future trajectory will not cover it. The affected trajectory is replaced by an unaffected trajectory, which is referred to as the trajectory replacement strategy. Here, the replacement methodology is elaborated for each restriction policy.

\noindent \textbf{Telecommuting}.
To implement telecommuting, first mobile phone users' homes and workplaces are extracted from the raw trajectory. After that, the administrative districts of the workplaces are identified to support the policy at the administrative district level. Given a group of telecommuting districts and a corresponding time range, users whose workplaces are in the target districts are initially determined. Then, each affected user's status in the time range is determined on a day-by-day basis, that is, whether they go to work. For the days on which a user goes to work, the trajectory of that day is replaced with the home address (i.e., make a stay at home and work remotely). For the days on which the user does not go to work, no changes are made.

\noindent \textbf{Regional Lockdown}.
Given a group of lockdown regions and time ranges, the regions are first mapped to a set of mesh grids covering them. Then, the affected users are divided into two categories. For users who stayed in the restricted area at the beginning of the lockdown period, their location remains unchanged throughout this period. For users who visited the restricted area during the lockdown, the affected trajectories are replaced with the unaffected trajectories. Specifically, a historical trajectory database is built for the users. For the affected day, the user's one-day trajectory that did not cross the restricted area is randomly selected from the historical database as the replacement.

\noindent \textbf{Screening}. \label{sec:screeningmodel}
In our system, users are allowed to set temperature screening points at the grid level. Contrary to the other two mobility restriction policies, the selected grids are screened during the epidemic simulation stage. Specifically, given a set of screening points and time ranges, all people in the grid are detected at each timestamp. If a person is healthy or in an incubation/latent period, it is assumed that the probability of an abnormal temperature is 0. If a person is infected, the probability of abnormal temperature is 87.9\%, which is set according to the latest research on COVID-19\cite{guan2020clinical}. 
Once an infected person is detected, he/she is quarantined (i.e., subsequent trajectories are discontinued) to prevent infecting others. Without loss of generality, EpiMob can also support other screening types (e.g., the nucleic acid test to detect latent patients).

\section{System Architecture}
\begin{figure}[t]
	\centering
	\includegraphics[width=0.45\textwidth]{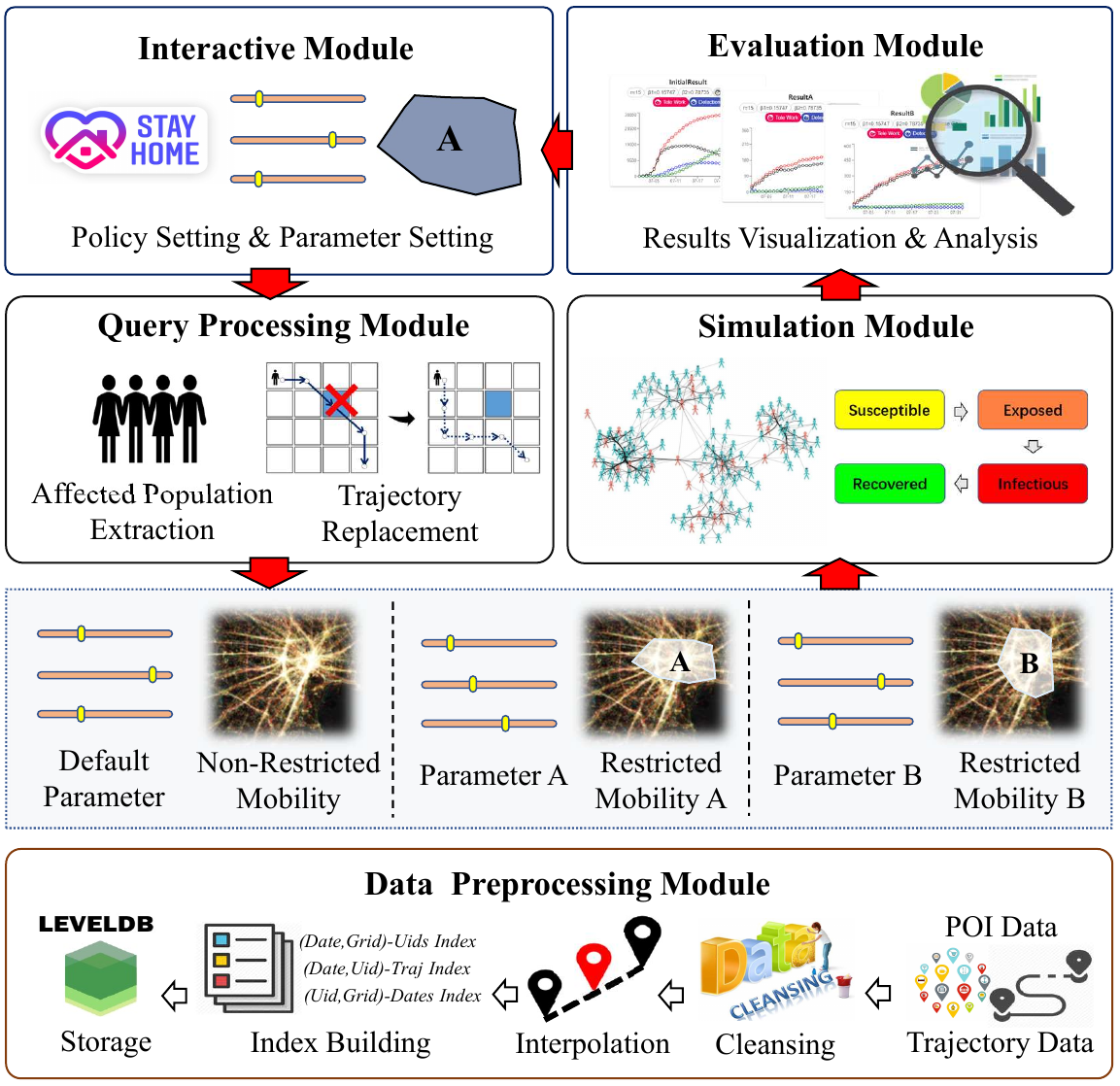}
	\caption{System Architecture of EpiMob comprises three parts: data storage, backend, and frontend. Data storage preprocesses and indexes the data in the database. Frontend helps users specify the epidemic parameters and control policies (the interactive module) and evaluates the visualization of the simulation results (the evaluation module). The backend interprets the control policies, generates a restricted trajectory dataset (the query-processing module), and simulates the spreading of the epidemic (the simulation module).}
	\label{fig:framework}
\end{figure}

EpiMob is a web application with a separate frontend and backend architecture. The frontend is implemented by React.js (for building user interfaces) and DECK.GL (for visual analysis of large-scale spatial data). The backend is designed as a Restful API implemented in Python. A set of integrated modules are utilized to construct the EpiMob system, as depicted in Fig. \ref{fig:framework}. 
\textit{The data preprocessing module} builds the data foundation of EpiMob. It performs data cleansing, interpolation, and indexing for raw citywide human mobility data and POI data and then stores them in LevelDB.
\textit{The interactive module} consists of an epidemic parameter setting view and three interactive policy setting views to help users specify the epidemic parameters and control policies.
\textit{The query processing module} receives policy settings from the interactive module. It first extracts the people affected by the given policy and then generates a substitution trajectory for each affected person. 
Given the restricted mobility dataset and epidemic parameters, \textit{the simulation module} simulates the spread of the epidemic.
\textit{The evaluation module} acquires, displays, and analyzes the results of the simulation. 

\section{Visual Design} 
To address the requirements identified in Section \ref{sec:taskanalysis}, for each view, we went through multiple iterations development and iteratively gathered feedback from experts, covering the visualization alternatives and interactive logic design. Finally, the visualization module is divided into three progressive submodules, corresponding to a complete manipulation cycle in EpiMob: the epidemic parameters of the disease are set first (C1), after which the control policies (E1, E2) are interactively set, and the results are displayed and analyzed (L1, L2).

\subsection{Epidemic Parameter Setting View}  \label{sec:basicsetingview}
This view is designed to assist users to flexibly specify epidemic parameters (C1). As shown in Fig. \ref{fig:teaser}-A4, there is a parameter settings panel provided for this purpose. The $\sigma$ and $\gamma$ are relatively simple to decide; there are two input boxes which enable users to enter their value directly. In terms of the dynamic $\beta$, the epidemic model (Section \ref{sec:core_model}) is designed with an input box to allow the user to set the value of $\beta_{global}$. The POI risk adjustment panel (Fig. \ref{fig:teaser}-A5) is for setting the risk value of each type of POI ($r_{i}$) and the scale adjustment factor $k$.
To make the risk setting both convenient and reasonable, three POI categories highly relevant to epidemic simulation are retained: entertainment place--\textit{high possibility for close contact}, restaurant--\textit{masks removed when eating}, and supermarket and shopping malls --\textit{big, confined spaces} after discussions with the experts. 
Selection of simulation periods is also supported. The user has the flexibility to choose the time range of the data they wish to include in the simulation, which helps experts explore the impact of periodic changes in human behavior on epidemic control.

\subsection{Spatiotemporal Restriction Setting View}\label{sec:interactivesettingview}
In this section, three interactive views of the restriction settings are introduced for E1 and E2. The view of each setting includes a spatial view displayed on a map (E1) and a setting panel (E2). The switches located in the upper-right corner of each panel (Fig. \ref{fig:teaser}-A1, A2, A3) set the visibility of corresponding spatial views, thus users can obtain sufficient and effective prior knowledge.

\subsubsection{Regional Lockdown View}
\begin{figure}[ht]
	\centering
	\includegraphics[width=0.45\textwidth]{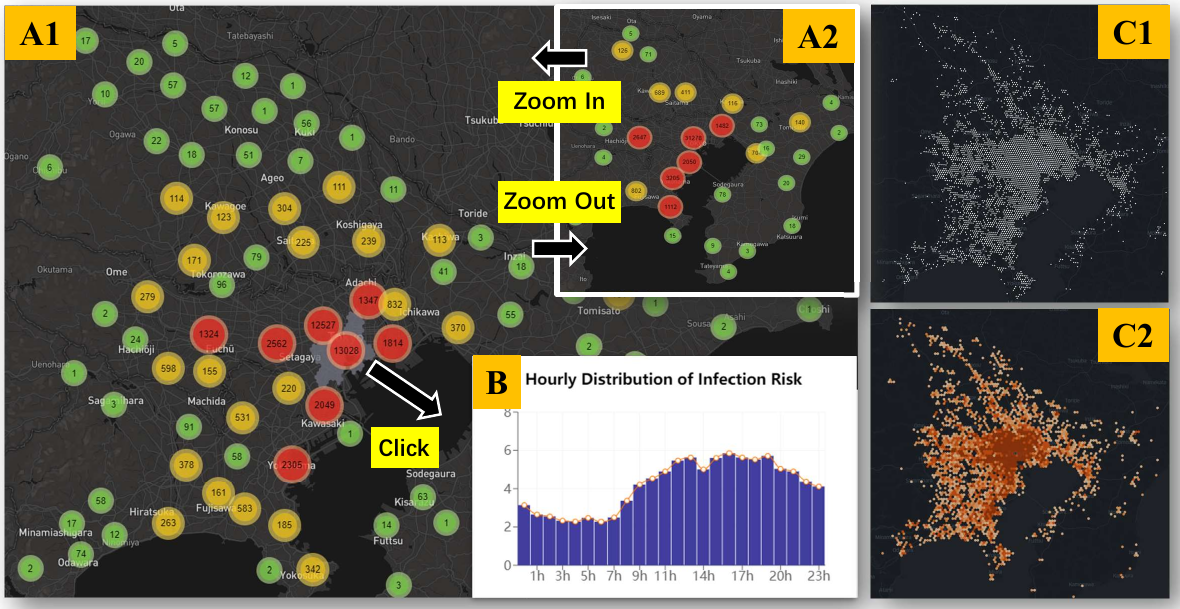}
	\caption{(A) shows the regional infection severity of the Greater Tokyo Area under no policy intervention at different spatial scales. Each marker indicates a cluster of infection events on that spatial scale. When the mouse hovers on a marker, the polygon that emerges denotes the spatial coverage range, and the number expresses infection severity. The color range from green to red reflects the infection severity from low to high. (B) By clicking the marker, hourly distribution of infection events are displayed. (C) shows two visualization alternatives: scatter plot (C1) and heatmap (C2).}
	\label{fig:riskmap1}
\end{figure}
\noindent To help users formulate appropriate lockdown policies, we devised the regional lockdown view, which consists of a map view showing the regional infection severity (Fig. \ref{fig:riskmap1}) and a setting panel (Fig. \ref{fig:teaser}-A1) allowing the corresponding parameters to be specified. Here, a definition of \textbf{regional infection severity} is provided---Given a region $R$ and a simulation result $E_{sim}$, the infection severity of $R$ indicates the accumulated number of infection events that occurred. The regional infection severity was defined as a reference indicator to determine the hotspots. To visualize the regional infection severity, specifying the simulation result $E_{sim}$ is required. A select box is placed---``Bind with'' in the setting panel (Fig. \ref{fig:teaser}-A1), where all existing simulation results are listed to allow users to choose. When using the regional lockdown view for the first time, it is recommended to perform a no policy simulation and then bind it to explore the infection hotspots under no intervention.

The regional infection severity view, derived from the marker cluster \cite{markercluster} view, helps visualize the regional infection severity at a multispatial scale (Fig. \ref{fig:riskmap1}-A). Markers represent spatial clusters of infection events at that spatial scale. As the mouse hovers on it, the spatial coverage is displayed on the map (A1). The number indicates the infection severity of the region covered. The greater the number, the more severe the infection is. In addition, the number is mapped to the marker's color to convey the severity (from green to orange to red). The spatial scale switches when performing zoom in/out, and the markers cluster together/spread out (A1, A2). Thus, users can observe infection hotspots at different spatial scales to formulate different policies. The scatter plot (C1) was discarded for this task because users could not recognize the hotspots within it. 
With the assistance of prior knowledge supplied by this view, the users can use the mouse to select the regions (i.e., polygons) they want to place under lockdown. For each selected region, EpiMob generates a configuration row in the setting panel for users to specify the start date and duration of the lockdown.

The elements of infection hotspots are infection events. The markercluster view is more suitable for presenting events than a heatmap. In addition to location information, events have other properties, such as the time of occurrence. The marker cluster makes it easy to aggregate and analyze events at multiple spatial scales by merely correlating the analysis with the marker's eventlistener (e.g., mouse click). In our system, the experts indicated that it is useful to analyze the temporal distribution of the infection events in a cluster to identify periods of high infection severity. Thus, a mouse click event was added to the marker. When the mouse is clicked, the hourly distribution of infection events is displayed in the form of a histogram (Fig. \ref{fig:riskmap1}-B). However, for the heatmap (Fig. \ref{fig:riskmap1}-C2), in-depth interaction analysis is not easy to integrate, even though it is also capable of finding the hotspots.

\subsubsection{Telecommuting View}
\begin{figure}[ht]
	\centering
	\includegraphics[width=0.48\textwidth]{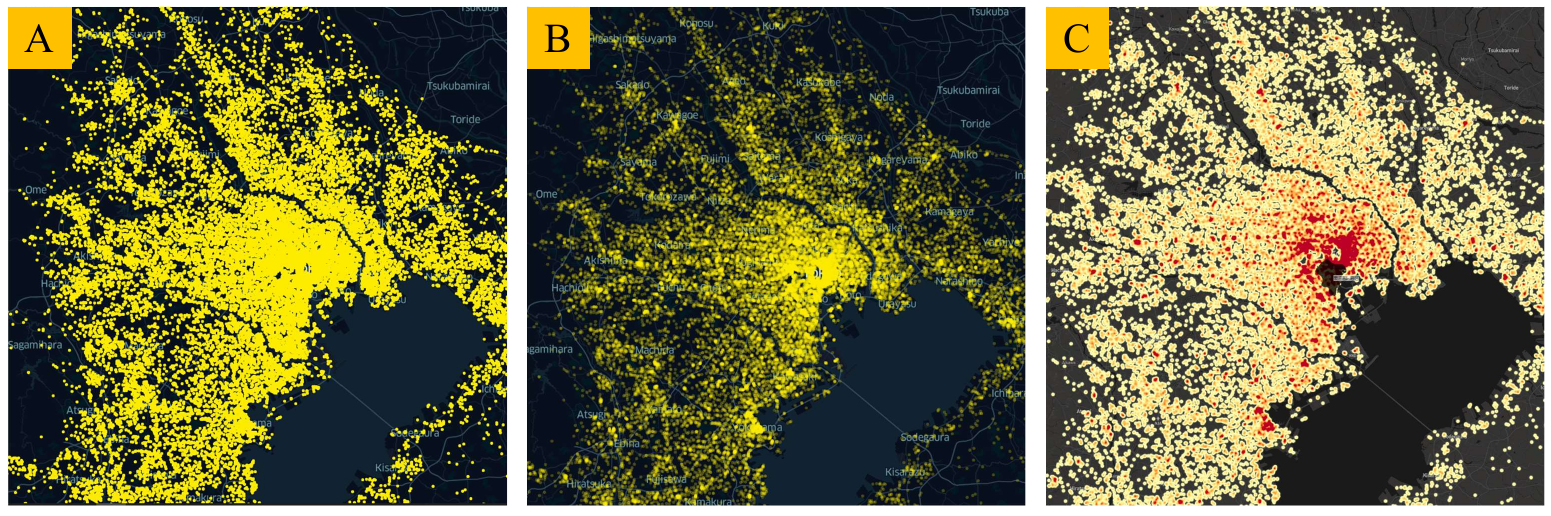}
	\caption{Workplace heatmap of the Greater Tokyo Area (C). The color range from yellow to red indicates the increasing number of people working there. The name of the corresponding administrative district is displayed when the mouse hovers over a region. A and B are design alternatives expressed with scatter plots in different point transparencies.}
	\label{fig:heatmap}
\end{figure}

To help determine remote work policies, a telecommuting view was designed, including a spatial view to identify workplaces that are frequented more regularly, along with a setting panel to finish the configuration of the concrete policy. Fig. \ref{fig:heatmap}-C shows the spatial view that displays all users' workplaces with a heatmap. The darker the color (i.e., red in this case), the more people work there. Because workplaces tend to have a high degree of overlap, for example, many people working in a single building, a heatmap was selected to depict the distribution of workplaces. Compared to scatter plots (Fig. \ref{fig:heatmap}-A\&B), the heatmap depicts the spatial accumulation of points more effectively. Moreover, when the mouse hovers over the heatmap, the name of the administrative district of the area is displayed. With this information, users can obtain the names of the regions to implement telecommuting.
Fig. \ref{fig:teaser}-A3 shows the panel for setting the telecommuting view. Users can set a series of regions to execute telecommuting here. This panel enables the ``reduction rate'' of regions to be specified, i.e., the percentage of people working from home in the target region, to control the intensity of execution. For example, ``Region Toshima Reduction 70\%'' means that 70\% of people working in Toshima are working from home. Moreover, according to regional conditions, the strength of the policy execution can also be quantified by setting the start date and duration. 

\subsubsection{Screening View}
\begin{figure}[ht]
	\centering
	\includegraphics[width=0.45\textwidth]{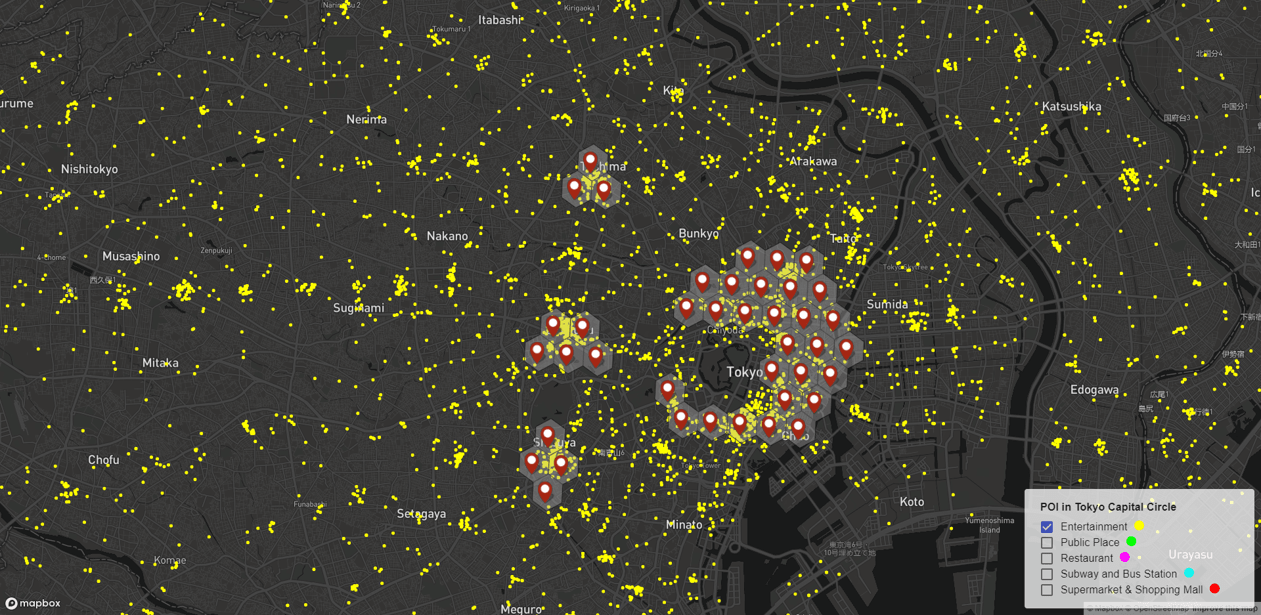}
	\caption{Case of the screening points setting. The user selects the locations with a denser distribution of entertainment places as screening points, as indicated by the red markers. The yellow scatter plot shows the spatial distribution of entertainment POIs. During policy implementation, all persons passing the marker-covering regions accept a temperature check. Users can simultaneously stack multiple types of POIs.}
	\label{fig:screening_view}
\end{figure}

A superimposable scatter plot is integrated to display POI information (Fig. \ref{fig:screening_view}). Users can select one or several types of POIs simultaneously to explore potential screening points. To achieve E2-iii, the user is allowed to set up while exploring. When the user finds a target region, they can draw selection areas directly on the map or drag markers to the locations of interest, as shown in the screening control panel (Fig. \ref{fig:teaser}-A2). After successfully adding a screening point, a mark is generated on the selected grid, indicating that screening will be performed. The time range for policy implementation can also be set. In the subsequent simulation, all people passing the screening points during the implementation of the policy are screened.

\subsection{Simulation Result View}\label{sec:solutionview}
This view was designed to assist users to intuitively observe the simulation results and conduct a comparative analysis (L1, L2), including two subviews: a single policy result view (L1) and a comparative analysis view (L2). Furthermore, a single policy result view consists of an Infection Curve (L1-i) and a Spatial Propagation Feature View (L1-iii).

\begin{figure*}[!t]
\centering
\includegraphics[width=0.9\textwidth]{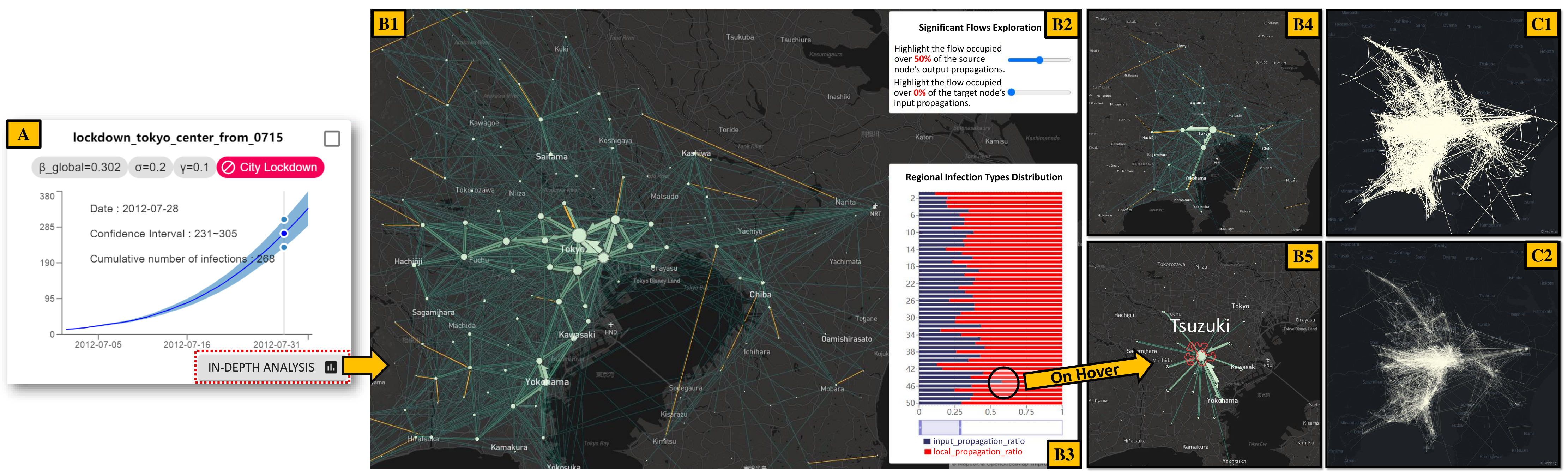}
\caption{ The result view for imposing lockdown on the central area of Tokyo since July 15th. (A) Cumulative infection curve. The purple area around the curve represents the 95\% CI. As the mouse hovers, it triggers a hover box, displaying the infection details of a certain date. The clips indicate the essential epidemic parameters and the type of policy implemented. (B) Spatial Propagation Feature View. B1 and B4 reveal the spatial transmission network on different zoom levels. B2 allows the user to set criteria to highlight propagation patterns (flows) of interest. In this case, the flows occupied over half of the source area's total output are highlighted in yellow. B3 enables users to track local infection patterns of interest. Here, the region having the highest input propagation is determined and displayed on the map (B5). (C) C1 and C2 are two spatial line chart alternatives for B1 with different line opacity settings.}
\label{fig:single_view}
\end{figure*}

\noindent \textbf{Infection Curve}.
The cumulative number of infections is used as the evaluation indicator; plots of the confidence intervals demonstrate the uncertainty, putting the naming initiative in the hands of users to help distinguish different policies (L1-ii); and adding clips to present details of the policy setting. Fig. \ref{fig:single_view}-A shows an instance of this view, named \textit{``lockdown\_tokyo\_center\_from\_20120715,''} referring to imposing the lockdown policy in central Tokyo since July 15th. The blue curve represents the cumulative number of infections, and the purple area represents the 95\% CI. The clips under the title represent the basic parameters and restriction types of the policy. The button in the lower right corner allows the user to explore the result in depth (Fig. \ref{fig:single_view}-B)
The checkbox in the upper-right corner of the view was designed to enable a comparative analysis. The analysis can be conducted by using the checkbox to first select the target policies and then clicking the compare button (bottom right corner of Fig. \ref{fig:teaser}-B). The corresponding results are displayed in the comparative analysis view, which combines multiple curves for comparative analysis (Fig. \ref{fig:teaser}-C). Similarly, the user can customize the name of the analysis result, and the evaluation indicator is the cumulative number of infections.

\noindent \textbf{Spatial Propagation Feature View}. 
This view (Fig. \ref{fig:single_view}-B) was designed to analyze the policies' secondary effects from the perspective of spatial propagation, and to understand the roles and patterns of different regions, including a map component (B1) to display \textit{the spatial transmission network}, a filter component (B2) to explore \textit{the significant cross-region propagation patterns}, and a tracing panel (B3) to local \textit{the local infection patterns of interest}. 
To this end, the ``initial infection location, secondary propagation location`` pairs of infected individuals were captured and treated as the subject of analysis. It is easy to find that the data conforms to origin and destination (OD) data characteristics.
A spatial line chart (Fig. \ref{fig:single_view}-C1) was first employed to visualize all OD pairs to display \textit{the spatial transmission network}, but the clutter problem is severe. The transparency of the line (C2) was further reduced, but the problem remained. One possible solution is OD aggregation visualization, but the clutter still exists according to \cite{phdthesis}. 
The final solution is presented in Fig. \ref{fig:single_view}-B, which inherits from a third party library \cite{flowmapview} designed for flow visualization. The bidirection of the edges encodes the direction of spatial diffusion. Both the width and color of the edges map the intensity of the transition. This library solves the clutter by stacking and highlighting the edges according to their transition intensity. It also supports scaling and clustering when the zoom level changes (B4). To trace \textit{the local infection patterns of interest}, all infections at the nodes were counted and categorized by infection type: \textit{input propagation}--the infection source comes from outside the area, and \textit{internal/local propagation}--the infection source comes from within the area. The local infection pattern is encoded as the respective proportion of the two infection types in the total infection (blue and red bar) to facilitate exploration and comparison. 
Finally, the infection distribution bars are listed in the order of node infection count (B3). When the mouse hovers on the bar of interest, its in-out flow, and spatial coverage are highlighted on the map. 
In this case, the area with the highest input propagation ratio in the top 50 infection count nodes was highlighted and it was determined that the infection mainly originated in Higashi-Kanagawa (B5). To explore \textit{the significant cross-region propagation patterns}, a filter panel is provided to identify the primary risk sources and output destinations of nodes (B2). The user is allowed to filter based on the proportion of flow to the total input in the target node and on the proportion of flow to the total output in the source node. The flows of nodes that meet the criteria are highlighted (in yellow in this study).

\section{Evaluation} \label{sec:eva}
In this section, case studies and expert interviews are presented to demonstrate the practicality and effectiveness of EpiMob. Here, descriptions of epidemic parameter settings and datasets are initially provided. 
\begin{itemize}[leftmargin=*]
    \item Through consultations with experts, the essential parameters of COVID-19 were determined from \cite{linton2020incubation}: $\sigma=0.2$, $\gamma=0.1$. $\beta_{global}$ was set to $0.302$ for a unit time of one day, as estimated by the fitting method proposed in \cite{kuniya2020prediction}.
    \item For POI-related settings, referring to the POI risk analysis section of \cite{chang2020mobility}, considering the real situation in Japan, and discussing with experts, the risk value of ``Entertainment Place,'' ``Restaurant,'' ``Supermarket and Shopping Mall'' were set to 8,2,1 respectively, and the scale adjustment factor $k$ was set to 0.0003.  
    \item The raw GPS trajectories of 30,000 mobile users collected for a period of one month (July 1 to 31, 2012) was pre-uploaded to EpiMob, randomly selected from the dataset of the Greater Tokyo Area. The simulation time range was from July 1st to 31st, 2012.
\end{itemize}

\subsection{Case Study}
Before conducting the case studies, a training seminar (30 min) was arranged to introduce the interactive logic of EpiMob. Subsequently, the experts worked together to evaluate the system. This section presents the evaluation process in several case studies, which can be summarized as follows: (i) to launch a no policy simulation to analyze the potential infection hotspots when COVID-19 spreads; (ii) to explore, evaluate, and analyze the effects of each type of control strategy and compound strategy, separately; and (iii) to perform in-depth analysis to explore the secondary effects after the implementation of the policies.
These procedures enabled the experts to gain helpful insights into potential epidemic prevention and control. Their discussion during the process was recorded. Finally, informal interviews were conducted for the feedback and suggestions of the experts.

\subsubsection{Epidemic Hotspots Exploration}
\begin{figure}[ht]
	\centering
	\includegraphics[width=0.45\textwidth]{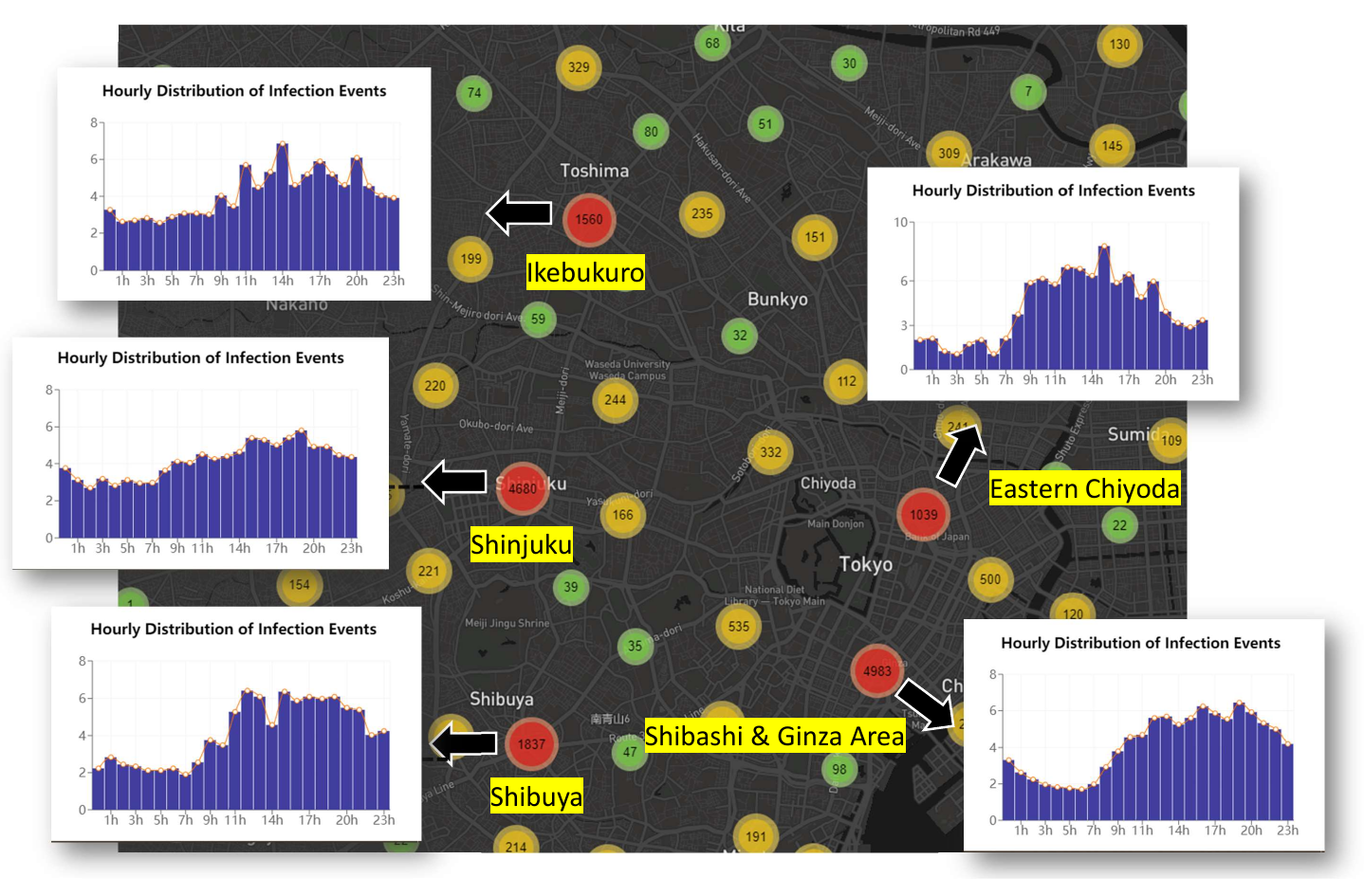}
	\caption{Top five infection hotspots in Tokyo under no policy intervention. The number of infection events in each of these hotspots exceeded 1,000, which was much higher than that of the surrounding areas. The histogram shows the hourly distribution of the infection events, revealing the infection time pattern in these areas.}
	\label{fig:case1}
\end{figure}
\noindent One of the experts wished to explore the potential hotspots of infection. He launched a basic no policy epidemic simulation with our system and bound it with a regional lockdown view. He discovered that most of the infections occurred in the central part of Tokyo (Fig. \ref{fig:case1}), distributed in five infection hotspots. All of the infection numbers in these areas were over 1,000. This finding is reasonable because these regions are well-known commercial, entertainment, and office areas in Japan, with an extremely high number of close contacts. In 2020, the news reported clusters of infections in these places. Furthermore, to explore the temporal patterns of infections in these areas, he successively clicked the corresponding markers to obtain the hourly infection distribution. He found that Shinjuku had the highest number of infections during the night (00:00–-06:00), and most of the infections in the Eastern Chiyoda occurred during work hours (11:00-17:00). This is because Shinjuku is known as Japan's largest city that never sleeps, and Eastern Chiyoda's day/night population ratio is the highest of all municipalities in Japan. Moreover, in all these five areas part of the infections occurred at nighttime, despite few people living in these places according to the census data. Thus, based on the situation explored, he concluded that the government could shorten the nighttime business hours in these areas, especially in Shinjuku, to prevent the infection cluster at night. The news confirmed his conclusion \cite{nighttimecloseall,nighttimecloseshijuku}.

\subsubsection{Epidemic Simulation under Control Policies} \label{sec:case_2}

\begin{figure*}[!t]
	\centering
	\includegraphics[width=0.9\textwidth]{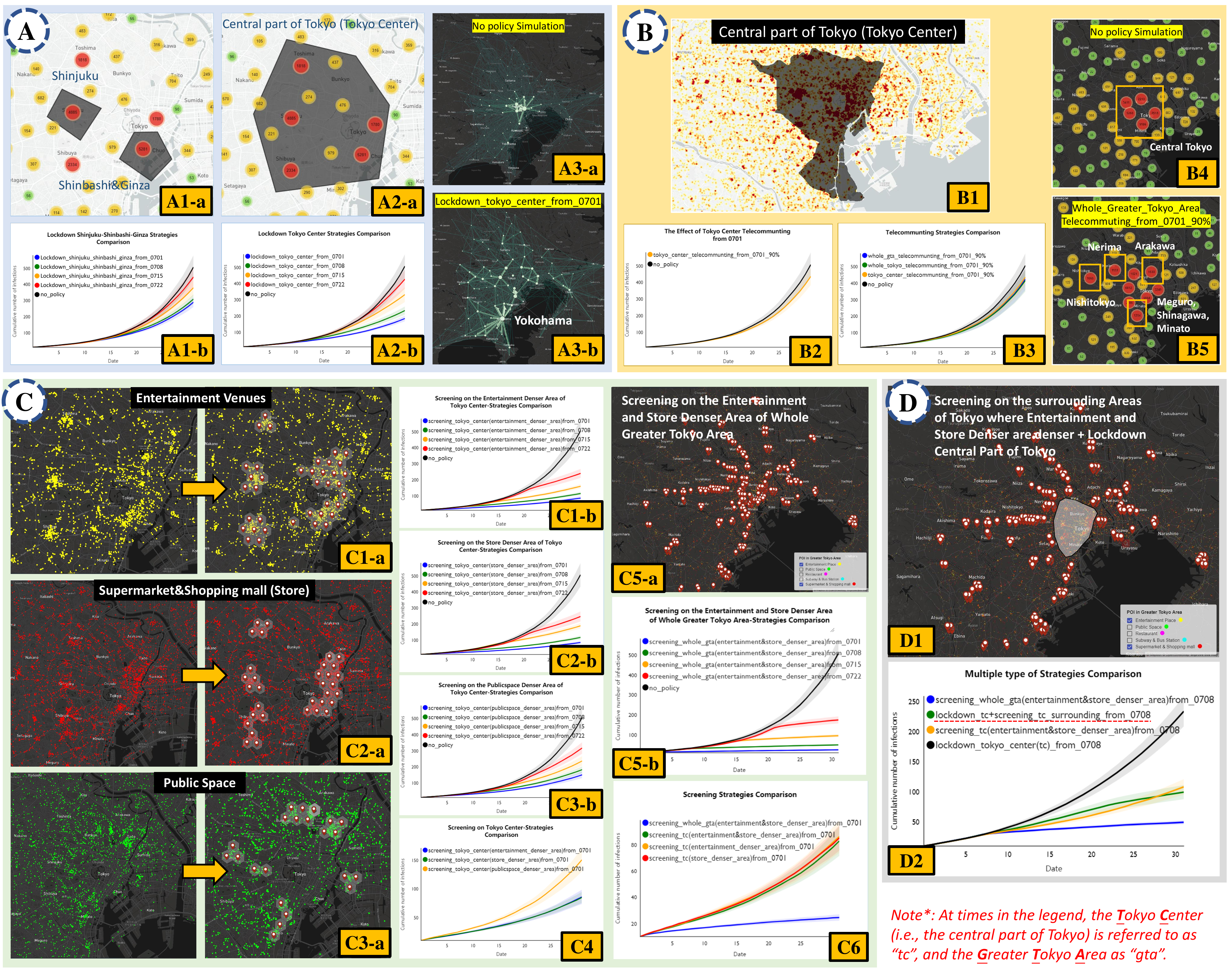}
	\caption{Comparison of epidemic results under restriction policies: (A) different lockdown policies; (B) different telecommuting policies; (C) various screening strategies; (D) compound policies.}
	\label{fig:casestudy}
\end{figure*}

\noindent\textbf{Regional Lockdown Policy [Fig. \ref{fig:casestudy}-A].} The experts ran a no policy simulation and obtained the regional lockdown view of the Greater Tokyo Area. They found that the central part of Tokyo clearly had a higher infection severity than other areas, especially in ``Shinjuku'' and ``Shinbashi\&Ginza Area'' (the number of infection events were around 5,000). 
Thus, they manually locked these two areas (A1-a) and launched a series of simulations with different durations of lockdown (from July 1st, 8th, 15th, 22rd until July 31st, separately). They found that the cumulative number of infections was reduced compared with that of no policy intervention (A1-b), and the effect diminished as the lockdown time was postponed. Then, they expanded the lockdown area to the central part of Tokyo (A2-a) and kept the other settings unchanged. Compared to the previous results, the spreading was further mitigated (A2-b). However, the gain in expanding the lockdown area becomes lower as the start time delay, \textit{``displays the property of diminishing marginal returns.''}
Furthermore, by observing the trend of the curves, they found that none of the partial lockdown policies could cut down the spread of the disease. They commented \textit{``Partial lockdown policies can just slow down the outbreak because some infected people carried the disease to other areas before,''} which corroborates the newest research \cite{lau2020positive,signorelli2020covid}. A preliminary in-depth analysis of the results also confirmed this. The spatial transition features with or without the lockdown policy in the central Tokyo area are shown in A3, where it can be observed that if the lockdown policy were to be applied, Yokohama, which is the second largest city next to Tokyo, would become the next ``epicenter'' of the infection spread. 

\noindent\textbf{Telecommuting Policy [Fig. \ref{fig:casestudy}-B].} The experts wanted to identify high-frequency workplaces for the purpose of imposing the telecommuting policy. Using the telecommuting view, they obtained a heatmap of the workplaces. They found that the central part of Tokyo is the most frequented workplace compared to other areas. Thus, they tried to implement a ``strict'' telecommuting policy (B1), which specifies the percentage of working from home people in this area as 90\% from the first day of simulation (working remotely is not possible for all occupations). They found that such a ``strict'' policy has only a slight effect (B2). Therefore, keeping the other settings unchanged, they expanded the telecommuting area to Tokyo and the entire Greater Tokyo Area. However, these efforts were still not effective enough (B3). An infection hotspot view of the results answered their doubts (B4), where they observed a shift of the hotspots from the central Tokyo area (B4) to the residential areas in Tokyo (B5). Based on the results, they concluded that: \textit{``The reason of the poor effect may be that the extreme telecommuting pushes more people to stay at home, but they are still free to move around on non-working days, thus leading to the increased clustering of household cases.''} The latest research \cite{prem2020effect} and news \cite{teleriskpush} corroborate the conclusions. 

\begin{figure}[!t]
	\centering
	\includegraphics[width=0.45\textwidth]{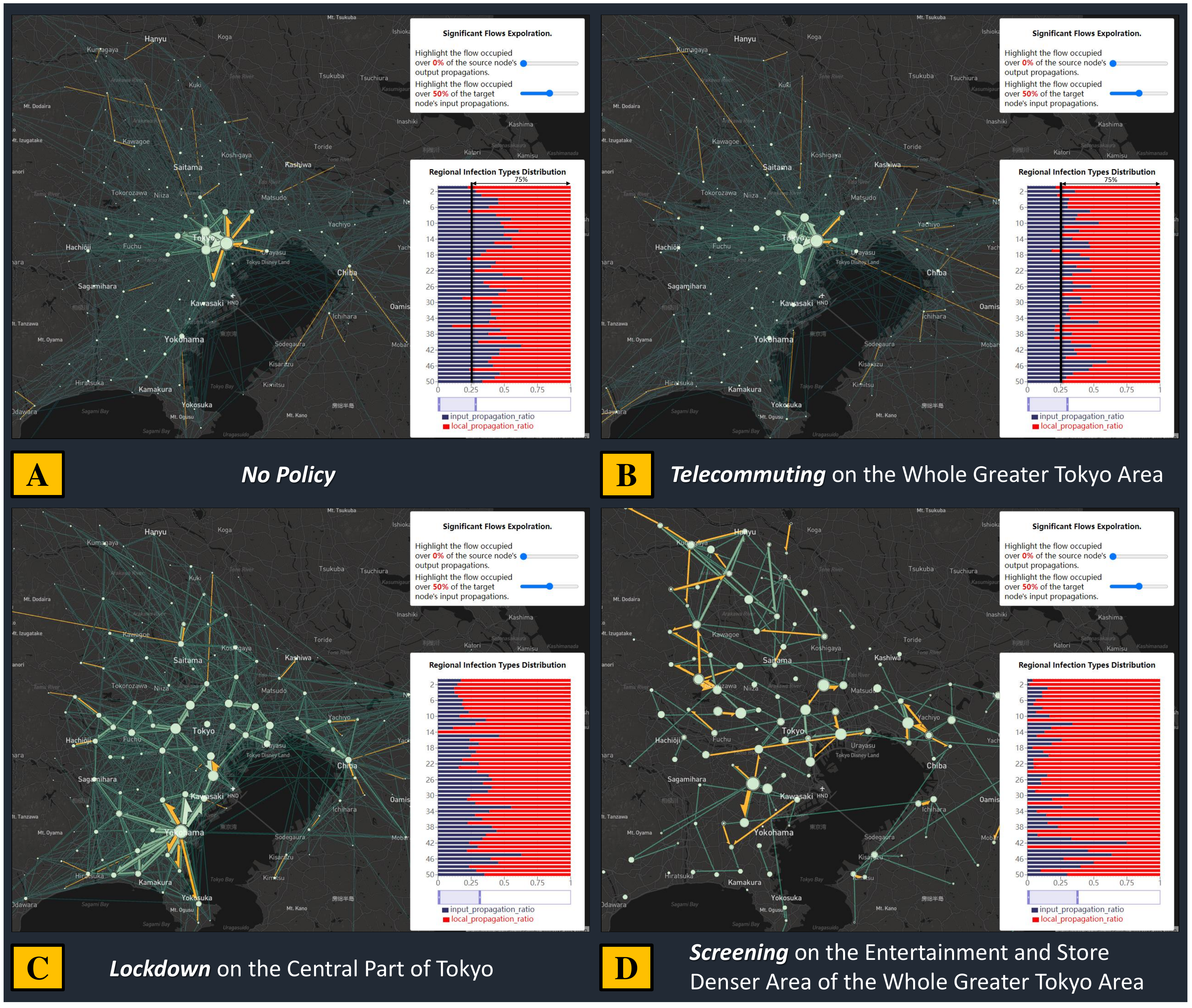}
	\caption{Spatial propagation features of a series of epidemic simulation results under different control policies; all of these policies took effect starting the first day of the simulation. For each result, the network view visually expresses the spatial transmission condition on the map, and the flow (i.e., the number of cross-region propagations) occupying over 50\% of the total input of the target node, highlighted in yellow to determine the significant risk sources. The bottom right panel shows the distribution of infection types for the top 50 infection severity regions. Each bar denotes a region, and the lengths of the red and blue sub-bars represent the ratio of internal infections and external import-caused infections, respectively.} 
	\label{fig:indepth_rr}
\end{figure}

\noindent\textbf{Screening Policy [Fig. \ref{fig:casestudy}-C].} The experts wanted to determine the appropriate screening points to impose the screening policy. They first opened the screening view and found that the central part of Tokyo is the densest area of the POI distribution compared with the surrounding areas. Thus, they initially attempted to set up screening points over there. They set up three groups of screening points at the places with denser entertainment venues, stores (i.e., supermarkets and shopping malls), and public places, respectively (C1-a, C2-a, C3-a). The spatial distribution of restaurants covers almost the entire central Tokyo area densely, and that of subway and bus stations is too sparse. Thus, these were not chosen as references. Subsequently, four simulations with different durations of screening were conducted for each group of selected screening points, and the results are displayed in C1-b, C2-b, and C3-b, separately. They noticed that all three screening strategies worked better than the previous lockdown and telecommuting results. 
Furthermore, by comparing the three screening strategies, they discovered that setting up temperature screening points in the denser areas of public places was less effective than those on entertainment venues or stores (C4). 
After that, keeping the other settings unchanged, they stacked the points of entertainment venues and stores together, and selected all the denser areas of these two POI types as the screening points. C6 shows the corresponding results. However, there was nearly no improvement because there were many overlaps between the denser areas of these two POI types. 
To further flatten the curve, they expanded the screening strategy to the entire Greater Tokyo Area (C5-a). They found that the curves flattened quickly, maintaining a slightly rising trend (C5-b), which provided the best group of results among all existing strategies (C6). According to the results, they concluded that:\textit{``In the Greater Tokyo Area, screening plus isolation is a very effective strategy, especially the large scale screening, because it cuts off the source of the infections.''} The latest research \cite{johanna2020mass,taipale2020population} confirms these observations.

\noindent\textbf{Compound Policy [Fig. \ref{fig:casestudy}-D].}
The experts finally tried to implement and analyze a combination of multiple policies. They locked down the \underline{T}okyo's \underline{c}entral part (abbreviated as ``tc'') and set up screening points at the locations where the \underline{e}ntertainment venues and \underline{s}tores (abbreviated as ``es'') are denser in the areas surrounding Tokyo (D1). As a control group, they selected three of the existing simulation results to compare, including: lockdown tc; screening the es denser area of tc; setting up screening on the es denser area of the whole \underline{G}reater \underline{T}okyo \underline{A}rea (abbreviated as "gta"). Furthermore, all above four policies were launched starting July 8th. Because (i) it is not realistic to launch a restriction policy from the first day of spreading; the government requires time to respond. (ii) As observed in Fig. \ref{fig:casestudy}, for two control policies with the same restriction type and spatial setting, the effect of launching from the first day of simulation (July 1st) and from one week later (July 8th) yielded similar results. 
D2 shows the final comparison result, temperature screening on the es denser area of whole gta is still the most effective. They discovered that relative to just screening on the es denser area of tc, D1 displayed almost no improvement, which highlights the importance of screening tc.

\subsubsection{Explore the Secondary Effects of Policies}
To understand the secondary impact of different policies, they conducted in-depth analysis of spatial propagation features under four different policies (Fig. \ref{fig:indepth_rr}). They also highlighted the significant flows that account for more than half of the total input propagation in a region to explore the major risk sources.
They found that under \textit{no policy simulation} (Fig. \ref{fig:indepth_rr}-A), the central part of Tokyo contributed significantly to the imported cases in the surrounding areas. More than half of the input propagation in many areas of the east came from the central part.
This effect diminished after \textit{telecommuting} was implemented (Fig. \ref{fig:indepth_rr}-B). Moreover, by comparing the local infection patterns of A and B, it is discovered that the extremes of the local propagation ratio have receded, and the areas with a high local propagation ratio (approximately 75\%) have become increasingly. Furthermore, experts separately hovered on these significant bars of interest (i.e., over or around 75\%) to locate the related regions. 
They found that the regions shifted from several core metropolitan areas (i.e., the central part of Tokyo, Chiba, and Yokohama) to the corresponding secondary regions around them (e.g., Kawasaki, Funabashi, and Kawaguchi), which further complements our conclusion regarding the effect of telecommuting.
The \textit{lockdown strategy} (Fig. \ref{fig:indepth_rr}-C) highlights the new ``epicenter,'' Yokohama, the transportation hub to the south of the Greater Tokyo Area, which could not be ignored if the Tokyo center were in lockdown. In addition, by observing the significant flows, they found that Yokohama contributed to more than half of the imported infections in various surrounding regions. Therefore, they suggested that the government should develop policies to prevent export risks from Yokohama, especially to the satellite cities surrounding it.
As the most effective strategy in the current simulation, they found that \textit{large-scale screening and isolation} in the Greater Tokyo Area (Fig. \ref{fig:indepth_rr}-D) broke the transmission network and scattered the infection cases to various locations: \textit{``rapid interception made it difficult to scale the cross-regional transmission.``} However, at the same time, the proportion of internal infections increased significantly; the input cases in the region were transferred from multiple sources to single sources; and the number of significant flows significantly increased. According to the results, they commented that: \textit{``tracing and isolating the small number of cases may become the new focus for prevention and control.``}

\subsection{Interviews with Domain Experts}
The feedback collected is summarized with respect to two aspects: visual design and simulation mechanism.

\noindent\textbf{Visualization Design.} All the experts appreciated the overall visualization design as \textit{``intuitive, practical, and easy to use, considerably simplified the process of policy modeling and search''}, and stated that the in-depth analysis \textit{``strengthened the judgment and understanding of how a policy works.''} They also provided some suggestions for improvement. EA commented that the view of the results \textit{``has only one indicator (the cumulative number of infections).''} He recommended adding more comparative indicators, such as \textit{``the R0 curve over time, which is more instructive for professionals.''} Moreover, the current UI does not allow a user-customized initial infection status (the number and corresponding locations of the initial infections). EC advised adding such a function for applicability to wider scenarios in the future.

\noindent\textbf{Simulation Mechanism.}
EB praised the conciseness of the simulation mechanism and the fact that it is practical and can be widely applied to other fields such as traffic control. Apart from this, the restriction strategies are more diverse in reality, \textit{``How about restricting the specific type of trips?''}. Adding such a function was recommend to assess the impact of different daily activities on the epidemic.

\section{Discussion}
\noindent \textbf{Lessons Learned}. Here, the two lessons learned during the design process are presented. (a) Visualization-assisted in-depth analysis makes the conclusions more comprehensive and persuasive. At the beginning of the project, the epidemic was urgent. The evaluation merely focused on observing and comparing the infection curves of the different strategies. Based on the curves, expertise, and experience, the experts gained insights into policy-making. Subsequently, the reviewers and experts expressed their expectations for an in-depth analysis. The subsequent results visually confirmed their previous judgments and helped them discover additional valuable insights. Compared with the plain curve, the new visual evidence is easier to understand and more shareable, receiving positive feedback from them. 
(b) The mainstream solutions for epidemic policy evaluation are still determined by manual modeling and static chart analysis, with much time spent on model design and coding. Similar to the reflections in \cite{Chen2021RAMPVISAT}, the common impression of the experts on visualization is still as a tool for information presentation or knowledge dissemination. Not only can we introduce our work within the VIS community, but we can also look for ways to increase the exposure of our work in the target audience.

\noindent \textbf{Implications}. Our work integrates a range of the latest visualization techniques and easy-to-use interaction logic to simplify cumbersome policy modeling, setting, and analysis, thus improving efficiency. It could change the common impression of visualization as a tool for information display or knowledge dissemination. Meanwhile, quality architectural design makes our system highly extensible. User-customized and contributed simulation models and control strategies further enhance the applicability of the system. For evaluation, the users explored the effectiveness and secondary impact of different policies and acquired insights, which are also helpful in the prevention and control of the pandemic in cities worldwide.

\noindent \textbf{Possible Directions and Challenges} for future visual analytics systems. To promote the design of future systems that feature epidemic simulation, additional interviews were conducted with experts to discuss how visualization can facilitate decision-making. EB maintains that the essence is \textit{``whether visualization can clearly and intuitively communicate what users care about most, to promote understanding and sharing.''} For epidemic control, perceiving the current status of the epidemic and determining the effectiveness and impact of implementing different strategies are the primary concerns of the decision-makers. A mechanism that can quickly and conveniently synchronize the current outbreak status is bound to enhance the applicability of the simulator, but the complexity of the outbreak status presents new challenges. It is also beneficial to display relevant contextual information (e.g., medical resources) when users need it.
EA further pointed out \textit{``policymakers will face new decision-making scenarios and goals as the epidemic evolves, e.g., contain the outbreaks/lift or introduce restrictions/zero the cases,``} that visualization should continually adapt to. They appreciated that the well-modularized architecture of EpiMob has great potential for scaling to include these new scenarios. However, due to human resource constraints and the urgency of the epidemic, collaboration with specialized development teams may be necessary for rapid requirements iteration and development. 
EC made further suggestions regarding policy implementation, arguing that facilitating decision-making can involve more than merely providing insights. It can also mean following the implementation of the policy in the real world to obtain real-time feedback, which will be the future for digital governance.

\noindent \textbf{Limitations and Future Work.}
\textit{(a) Visualization.} The information that EpiMob can provide and cover is limited, and not all experts and officials need to make decisions. Integrating more context information, such as medical resource status, will be helpful. \textit{(b) Simulation.} Our method is highly dependent on the quality of the trajectory data. If the trajectory data are very biased and cannot reflect the movement pattern of the city, the accuracy of the result may not be guaranteed. Moreover, designing an enhanced model that considers the scaling factor (mapping the limited trajectories to the real population) could make the model more evaluation friendly.
\textit{(c) Applicability.} When the target area is a cluster of dozens of cities, it is too complex to set the parameters one by one. An automated parameter calibration mechanism or a machine learning method for parameter migration may be required to enhance the applicability of the system. In summary, the focus, in the future, will be on enhancing our system in two aspects. For visualization, this would involve integrating more context information, adding more comparative indicators for expert users, and continuing to explore new candidates for existing views. For simulation, this would involve designing automated parameters, taking the scaling factor into consideration when designing the model and the infection status calibration mechanism to quickly adapt to different scenarios and to develop and support more epidemic control policies.

\section{Conclusion}
In this study, an interactive visual analytics system called EpiMob was designed to effectively measure and evaluate different human mobility restrictions for epidemic control. 
EpiMob enables users to easily select one policy or a combination of policies as a simulation target with interactive visual assistance. By employing advanced visualization techniques, the simulation results can be confirmed, compared, and deeply analyzed in a well-organized and user-friendly layout. The functionality and usability of our system were validated by conducting multiple case studies and interviews with domain experts. Even so, there are still some shortcomings, and we expect to address them in subsequent work.

\ifCLASSOPTIONcaptionsoff
  \newpage
\fi

\bibliographystyle{IEEEtran}
\bibliography{IEEEabrv,reference}

\begin{IEEEbiography}[{\includegraphics[width=1in,height=1.25in,clip,keepaspectratio]{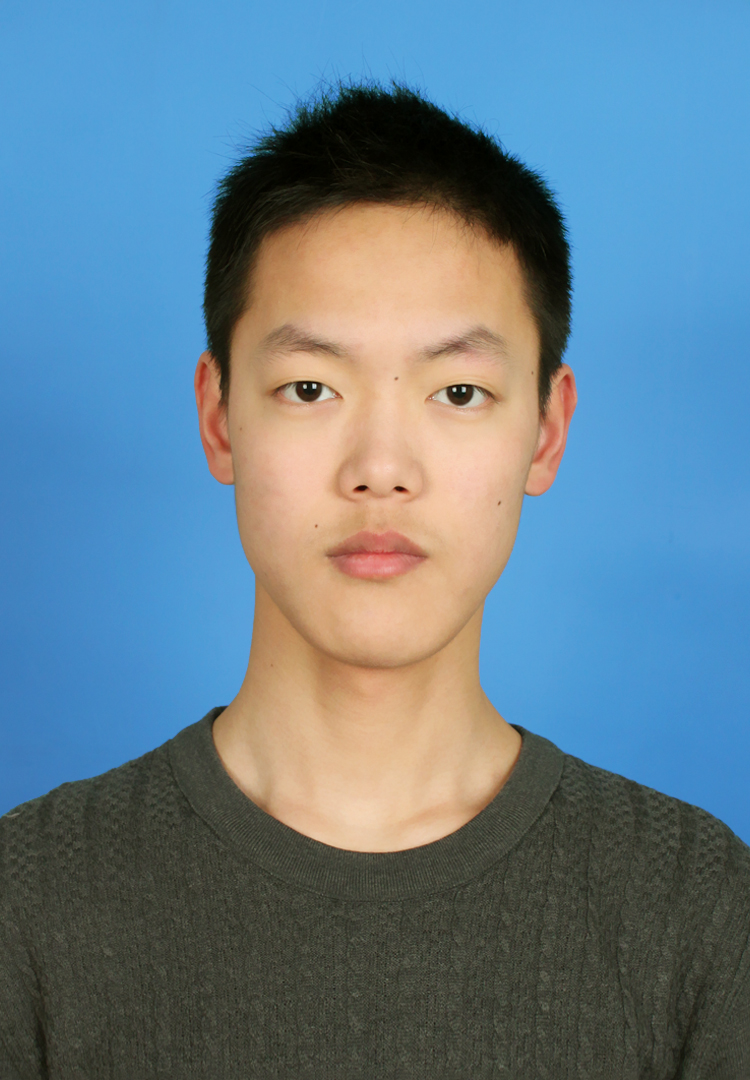}}]{Chuang Yang}
 received his BS degree in computer science and technology from Southern University of Science and Technology (SUSTech) in 2019. From 2019, he became a research student at the Center for Spatial Information Science, the University of Tokyo. His current research interests are in the area of spatiotemporal data analysis and mining, data visualization. 
\end{IEEEbiography}

\begin{IEEEbiography}[{\includegraphics[width=1in,height=1.25in,clip,keepaspectratio]{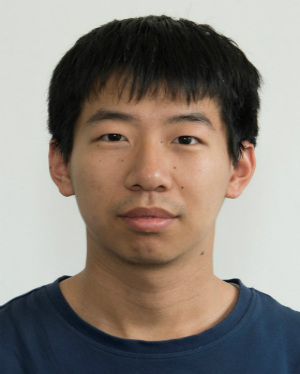}}]{Zhiwen Zhang}
received his BE and MS degrees in artificial intelligence from Nankai University, China, in 2016 and 2019, respectively. From 2019, he began pursing a PhD in the Department of Socio-Cultural Environmental Studies, the University of Tokyo. His current research interests include urban computing and data visualization.
\end{IEEEbiography}

\begin{IEEEbiography}[{\includegraphics[width=1in,height=1.25in,clip,keepaspectratio]{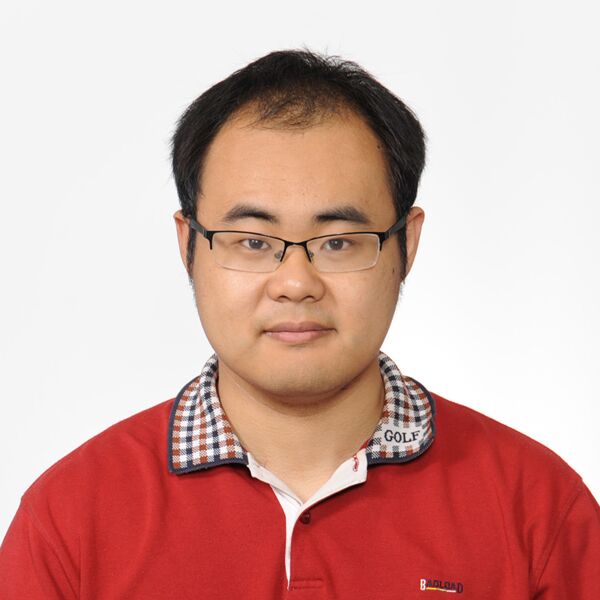}}]{Zipei Fan}
	received his BS degree in computer science from Beihang University, China, in 2012, his MS degree and PhD in civil engineering from the University of Tokyo, Japan, in 2014 and 2017, respectively. From 2017, he became a Project Assistant Professor at the Center for Spatial Information Science, the University of Tokyo. His research interests include ubiquitous computing, machine learning, spatiotemporal data mining, and heterogeneous data fusion.
\end{IEEEbiography}

\begin{IEEEbiography}[{\includegraphics[width=1in,height=1.25in,clip,keepaspectratio]{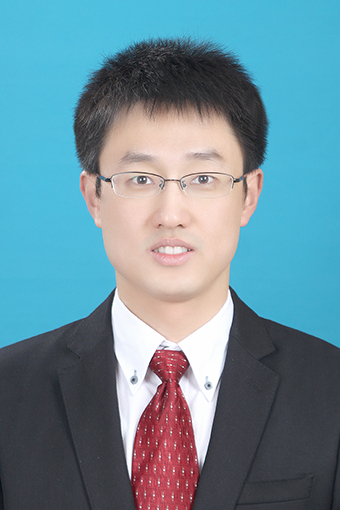}}]{Renhe Jiang}
	received his BS degree in software engineering from Dalian University of Technology, China, in 2012, MS degree in information science from Nagoya University, Japan, in 2015, and PhD in civil engineering from the University of Tokyo, Japan, in 2019. From 2019, he became an Assistant Professor at the Information Technology Center, the University of Tokyo. His research interests include ubiquitous computing, deep learning, and spatiotemporal data analysis.
\end{IEEEbiography}

\begin{IEEEbiography}[{\includegraphics[width=1in,height=1.25in,clip,keepaspectratio]{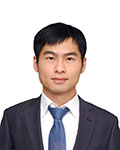}}]{Quanjun Chen}
	received his BS degree in automation from Hunan University, China, in 2011, his MS degree in pattern recognition and intelligent systems from Beihang University in 2014, and his PhD in civil engineering from the University of Tokyo, Japan in 2017. In 2017, he became a Post-Doctoral Researcher at the Center for Spatial Information Science, the University of Tokyo. His research interests include ubiquitous computing, machine learning, and traffic accident analysis.
\end{IEEEbiography}

\begin{IEEEbiography}[{\includegraphics[width=1in,height=1.25in,clip,keepaspectratio]{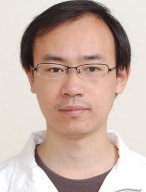}}]{Xuan Song}
	received his BS degree in information engineering from Jilin University, China, in 2005 and his PhD in signal and information processing from Peking University, China in 2010. He was promoted to Project Assistant Professor and Project Associate Professor at the Center for Spatial Information Science, the University of Tokyo, in 2012 and 2015, respectively. His research is mainly in the areas of artificial intelligence, computer vision, and robotics, particularly smart cities and intelligent system design.
\end{IEEEbiography}

\begin{IEEEbiography}[{\includegraphics[width=1in,height=1.25in,clip,keepaspectratio]{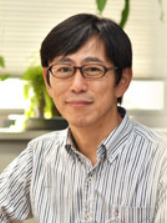}}]{Ryosuke Shibasaki}
was born in Fukuoka, Japan. He received his BS and MS degrees and is PhD in civil engineering from the University of Tokyo, Japan in 1980, 1982, and 1987, respectively. From 1982 to 1988, he was with the Public Works Research Institute, Ministry of Construction. From 1988 to 1991, he was an Associate Professor in the Civil Engineering Department, the University of Tokyo. In 1991, he joined the Institute of Industrial Science, the University of Tokyo. In 1998, he was promoted to Professor in the Center for Spatial Information Science, the University of Tokyo. His research interests cover three-dimensional data acquisition for GIS, conceptual modeling for spatial objects, and agent-based microsimulations in a GIS environment.
\end{IEEEbiography}

\end{document}